\documentclass[twocolumn,showpacs,preprintnumbers,amsmath,amssymb,aps,prb,longbibliography]{revtex4-2}
\usepackage{graphicx}
\usepackage{dcolumn}
\usepackage{bm}
\usepackage{verbatim} 
\usepackage[usenames, dvipsnames]{color} 
\usepackage{ulem}
\usepackage{color}
\usepackage[colorlinks=false,pdfencoding=auto,psdextra]{hyperref}
\usepackage{multirow}
\usepackage{tabularx}
\usepackage{dsfont}

\definecolor{orange}{rgb}{1.0, 0.5, 0.0}

\newcommand{\delV}[1]{\textcolor{blue}{}}

\begin{document}
\title{Optical spin orientation of localized electrons and holes interacting with nuclei in an FA$_{0.9}$Cs$_{0.1}$PbI$_{2.8}$Br$_{0.2}$ perovskite crystal}

\author{Dennis~Kudlacik$^{1}$, Nataliia~E.~Kopteva$^{1}$, Mladen~Kotur$^{1}$, Dmitri~R.~Yakovlev$^{1}$, Kirill V. Kavokin$^{2}$, Carolin~Harkort$^{1}$, Marek~Karzel$^{1}$,  Evgeny~A.~Zhukov$^{1}$, Eiko~Evers$^{1}$, Vasilii~V.~Belykh$^{1}$, and Manfred~Bayer$^{1}$} 

\affiliation{$^{1}$Experimentelle Physik 2, Technische Universit\"at Dortmund, 44227 Dortmund, Germany}
\affiliation{$^{2}$Spin Optics Laboratory, Saint Petersburg State University, 198504 Saint Petersburg, Russia}

\date{\today}
\makeatletter
\newenvironment{mywidetext}{%
  \par\ignorespaces
  \setbox\widetext@top\vbox{%
   \hb@xt@\hsize{%
    \leaders\hrule\hfil
    \vrule\@height6\p@
   }%
  }%
  \setbox\widetext@bot\hb@xt@\hsize{%
    \vrule\@depth6\p@
    \leaders\hrule\hfil
  }%
  \onecolumngrid
  \vskip10\p@
  \dimen@\ht\widetext@top\advance\dimen@\dp\widetext@top
  \cleaders\box\widetext@top\vskip\dimen@
  \vskip6\p@
  \prep@math@patch
}{%
  \par
  \vskip6\p@
  \setbox\widetext@bot\vbox{%
   \hb@xt@\hsize{\hfil\box\widetext@bot}%
  }%
  \dimen@\ht\widetext@bot\advance\dimen@\dp\widetext@bot
  \vskip\dimen@
  \vskip8.5\p@
  \twocolumngrid\global\@ignoretrue
  \@endpetrue
}%
\makeatother

\begin{abstract}
Optical orientation of carrier spins by circularly polarized light is the basic concept and tool of spin physics in semiconductors. We study the optical orientation of electrons and holes in a crystal of the FA$_{0.9}$Cs$_{0.1}$PbI$_{2.8}$Br$_{0.2}$ lead halide perovskite by means of polarized photoluminescence, time-resolved differential reflectivity, and time-resolved Kerr rotation. At the cryogenic temperature of 1.6~K the optical orientation degree measured for continuous-wave excitaton reaches 6\% for localized electrons and 2\% for localized holes. Their contributions are distinguished from each other and from exciton optical orientation through the pronounced Hanle effect in transverse magnetic fields and the polarization recovery effect in longitudinal magnetic fields. The optical orientation degree is highly stable against detuning of the laser photon energy from the band gap by up to 0.25~eV, showing then a gradual decrease for detunings up to 0.9~eV. This evidences the inefficiency of spin relaxation mechanisms for free carriers during their energy relaxation. Spin relaxation for localized electrons and holes is provided by the hyperfine interaction with the nuclear spins. Dynamic polarization of nuclear spins is demonstrated by the Overhauser field reaching 4~mT acting on the electrons and $-76$~mT  acting on the holes. This confirms the specifics of lead halide perovskite semiconductors, where the hole hyperfine interaction with the nuclei considerably exceeds that of the electron.  
\end{abstract}

\maketitle


Lead halide perovskite semiconductors are emerging materials attracting great interest due to their high photovoltaic efficiency and remarkable optoelectronic properties~\cite{Vinattieri2021_book,Vardeny2022_book}. Their band structure, which considerably  differs from conventional III-V and II-VI semiconductors, positions them as interesting materials for spin physics and for applications based on spin-dependent phenomena, e.g., in quantum information technologies~\cite{Vardeny2022_book,wang2019,kim2021}. The optical and magneto-optical techniques used for studying the spin physics of semiconductors~\cite{OO_book,Spin_book_2017}, like time-resolved Faraday/Kerr rotation~\cite{odenthal2017,belykh2019}, spin-flip Raman scattering~\cite{kirstein2022nc}, circularly polarized photoluminescence (PL), either induced by circularly polarized excitation~\cite{wang2019,Wang2018,wu2019} or by external magnetic field~\cite{zhang2015,zhang_field-induced_2018}, polarization-sensitive time-resolved differential transmission~\cite{Giovanni2015,zhou2020}, optically-detected magnetic resonance~\cite{kirstein2022am}, etc. are working well also for perovskite semiconductors.

Most of these techniques require an initial spin polarization of carriers or excitons, which is gained by their optical orientation using circularly polarized laser light. In perovskite semiconductors, the electrons at the bottom of the conduction band and the holes at the top of the valence bands have spin $1/2$, which allows one to reach a maximal spin orientation of carriers of up to 100\% right after their photogeneration~\cite{XOO}. During energy relaxation and before recombination, the degree of optical orientation decreases due to spin relaxation. In experiments on optical orientation, the circular polarization degree of the subsequent photoluminescence is measured as result of circularly polarized photoexcitation. In this case the ratio of the carrier or exciton lifetimes to their spin relaxation times determines the detected degree of optical orientation. 

Optical orientation measured under continuous-wave excitation was reported for polycrystalline films of MAPbBr$_3$  (optical orientation degree of 3.1\%~\cite{wang2019} at 10~K temperature, 2\%~\cite{Wang2018} and 8\%~\cite{wu2019} at 77~K) and of MAPbI$_3$ (0.15\%~\cite{Wang2018} at 77~K). These values are far below the ultimate limit of 100\%, evidencing very efficient spin relaxation, which could be also related to the carrier scattering at the polycrystalline boundaries. Recently, the giant optical orientation degree of 85\% was found for excitons at 1.6~K~\cite{XOO} in bulk crystals of FA$_{0.9}$Cs$_{0.1}$PbI$_{2.8}$Br$_{0.2}$ with a high structural quality. Here, the time-resolved PL technique was used to distinguish excitons with short lifetime of 60~ps from recombination of spatially separated electrons and holes with much longer lifetimes. In great contrast to conventional semiconductors with zinc-blend crystal structure, like GaAs or CdTe, the optical orientation degree in this sample was remarkably stable against detuning of the laser photon energy from the exciton resonance by up to 0.3~eV. It shows that the energy relaxation of excitons/carriers with large kinetic energies is not accoumpanied by their spin relaxation, which can be explained by the absence of the Dyakonov-Perel spin relaxation mechanism~\cite{OO_book,Spin_book_2017} in the centrosymmetric crystals of cubic lead halide perovskites.
\begin{figure*}[htb]
\begin{center}
\includegraphics[width = 18cm]{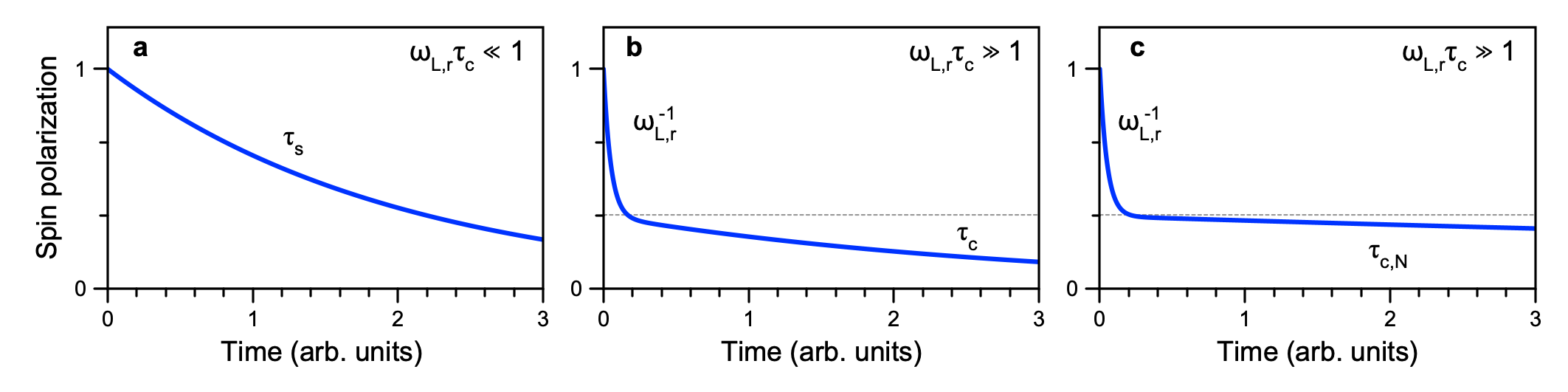}
\caption{\label{dynamics} Modeled regimes of the spin polarization dynamics with the spin relaxation governed by the interaction with nuclear spins. Regime of: (a) short correlation times $\omega_\text{L,r}\tau_\text{c} \ll 1$, (b) intermediate correlation times  $\omega_\text{L,r}\tau_\text{c} \gg 1$, $\tau_\text{c} < \tau_\text{c,N}$, and (c) infinitely long correlation times $\omega_\text{L,r}\tau_\text{c} \gg 1$, $\tau_\text{c} \gg \tau_\text{c,N}$.} 
\end{center}
\end{figure*}

In this paper, we study the same FA$_{0.9}$Cs$_{0.1}$PbI$_{2.8}$Br$_{0.2}$ perovskite crystals, for which the giant optical orientation of excitons was found, but focus on the optical orientation of localized electrons and holes as well as on the effects of their interaction with the nuclear spin system. Most of the experiments are performed using continuous-wave excitation and analyzing polarized emission.  Measurements in transverse (Hanle effect) and  longitudinal (polarization recovery effect) magnetic fields allow us to distinguish contributions of electrons, holes, and excitons. We show that at cryogenic temperatures the spin relaxation is provided by the interaction with nuclear spin fluctuations. In turn, the nuclear spins can be dynamically polarized by the spin-oriented carriers, with the holes showing larger efficiency due to their stronger hyperfine interaction compared to that of the electrons.  

\section{Introduction: optical spin orientation}

\begin{figure*}[htb]
\begin{center}
\includegraphics[width = 16cm]{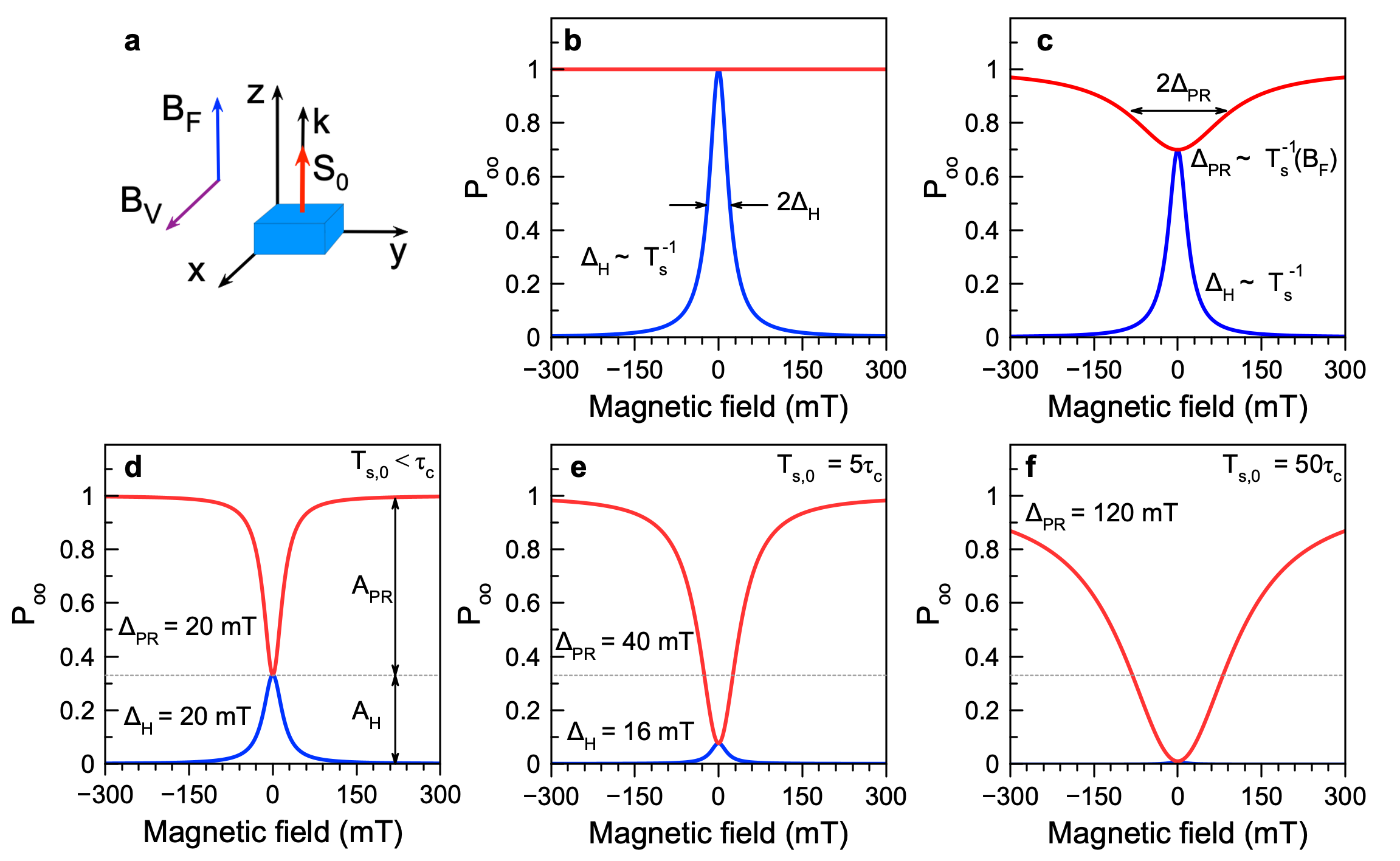}
\caption{\label{tc} Modeling of different regimes of the Hanle and PR effects. (a) Schematics of the experimental geometry. (b,c) Hanle (blue) and polarization recovery (red) curves for  negligible carrier-nuclear spin interactions for two cases: (b) spin relaxation time does not depend on $B_\text{F}$ and $\Delta_\text{H}\sim T_\text{s}^{-1}$ , and (c) $\tau_\text{s} = \tau_\text{s}(B_\text{F})$ and $\Delta_\text{H}\sim T_\text{s}^{-1}$. (d-f) Hanle (blue) and PR (red) curves for considerable carrier-nuclear spin interactions and for the intermediate regime of correlation time $\omega_\text{L,r}\tau_\text{c} \gg 1$, $\tau_\text{c}<\tau_\text{c,N}$. Three cases are modeled: $T_\text{s,0} < \tau_\text{c}$, $T_\text{s,0}  = 5 \tau_\text{c}$, and $T_\text{s,0} = 50 \tau_\text{c}$. Saturation of the optical orientation degree in longitudinal field $P_\text{max} = 1$ is assumed for simplicity. In experiment, it is lower due to the spin relaxation within the carrier recombination time.}  
\end{center}
\end{figure*}

Optical orientation in semiconductors is based on excitation of non-equilibrium electron-hole pairs with circularly polarized light, either continuous-wave or pulsed. Depending on the specific band structure, the angular momentum $\pm1$ of circularly polarized photons can be transferred to the photocreated carriers and distributed in a certain proportion between the electron and the hole. In the particular case of lead halide perovskites, the optical selection rules predict that absorption of right-circular polarized ($\sigma^+$) photons would result in creation of electrons and holes with 100\% spin polarization along the wave vector of light, due to spin $1/2$ for both carriers and the large spin-orbital splitting of the upper conduction band. This remains true over a wide range of photon energies above the band gap, see Ref.~\cite{XOO}. 

The lifetime of the spin-polarized state $T_\text{s}$ of electron (hole) is then limited either by the recombination time $\tau$ or the spin relaxation time $\tau_\text{s}$, according to $T_\text{s} = (\tau_\text{s}^{-1} + \tau^{-1})^{-1}$. Several spin relaxation mechanisms are known for semiconductors~\cite{OO_book,Spin_book_2017}. Most of them can be interpreted as result of electron (hole) spin precession in the random effective magnetic fields created by the spin-orbit and spin-spin interactions, including the hyperfine interaction with randomly oriented nuclear spins.

In III-V and II-VI bulk semiconductors and their quantum wells, the correlation time of these random fields, $\tau_\text{c}$, is so short that during this time the spin turns through by a very small angle, resulting in the relation $\omega_\text{L,r}\tau_\text{c} \ll 1$, where $\omega_\text{L,r}$ is Larmor frequency of the spin precession in the random field. 
In this regime spin relaxation occurs via many consequent small-angle rotations of the spin around randomly directed axes, which results in an exponential decay of the mean carrier spin $\langle S(t) \rangle$ with the spin relaxation time $\tau_\text{s}$ inversely proportional to the mean squared $\omega_\text{L,r}$ and $\tau_\text{c}$:
\begin{equation}
\label{eq_St}
\langle S(t) \rangle = \langle S(0) \rangle \exp(-t/\tau_\text{s}),
\end{equation}
\begin{equation}
\label{eq_tauS}
\tau_\text{s} = \left(\frac{2}{3}\langle \omega_\text{L,r}^2\rangle\tau_\text{c} \right)^{-1}.
\end{equation}
Here, $\langle S(0) \rangle$ is the initial spin polarization created by a light pulse. 

The spin relaxation in the regime of short correlation time $\omega_\text{L,r}\tau_\text{c} \ll 1$ is relevant, e.g., for the Dyakonov-Perel mechanism (spin precession in the spin-orbit fields arising when an electron moves in a crystal lacking spatial inversion symmetry like GaAs, CdTe and ZnSe~\cite{OO_book_Ch2,OO_book_Ch3}), or for the hyperfine relaxation of donor-bound electrons in GaAs~\cite{Korenev}. In the latter case, the random fields are created by nuclear spin fluctuations around donors via their hyperfine interaction with the electron spin, and the correlation time $\tau_\text{c}$ is defined by electron hopping to other donors in the impurity band. The spin polarization dynamics in the regime of short correlation time is monoexponential with the spin relaxation time $\tau_\text{s}$, as illustrated in Figure~\ref{dynamics}a.

The opposite regime of infinitely long correlation time $\omega_\text{L,r}\tau_\text{c} \gg 1$  and $\tau_\text{c} \gg \tau_\text{c,N}$ is realized in quantum dots~\cite{Merkulov,Braun2005}. Here $\tau_\text{c,N}$ is the correlation time of the nuclear spin fluctuations. In this regime, the electron spin is subject to a quasi-static effective magnetic field created by randomly oriented spins of a large, but finite number of lattice nuclei in the quantum dot. As a result, the average electron spin in a quantum dot ensemble decays non-exponentially down to 1/3 of the initial value during a time of the order of $\omega_\text{L,r}^{-1}$. Thereafter, the electron spin in each quantum dot is projected to the local field direction, and then remains in this state. Since the nuclear spins slowly change their direction, the electron spin polarization in the quantum dot ensemble further decays within the correlation time of the nuclear spin fluctuations $\tau_\text{c,N}$.  This time, determined either by the hyperfine interaction with the electron or by the dipole-dipole interaction between neighbouring nuclear spins~\cite{Spin_book_2017_Ch12}, typically is in the microsecond range.  
The corresponding spin dynamics is shown in Figure~\ref{dynamics}c. 

The intermediate regime of a long but finite correlation time $\omega_\text{L,r}\tau_\text{c} \gg 1$ and $\tau_\text{c}<\tau_\text{c,N}$ is rarely realized in conventional III-V semiconductors (Ref.~\cite{Korenev} presents such an example). In this regime the electron (hole) spin also rapidly projects on the local nuclear spin fluctuation, but then the electron (hole) gets the chance to hop to another localization center with different orientation of the nuclear fluctuation field. As a result, after the rapid fall to 1/3 of the initial value, the mean spin continues to decay with a characteristic time close to $\tau_\text{c}$, as shown in Figure~\ref{dynamics}b. The full theoretical description of the spin dynamics in the intermediate regime is given in the Supporting Information, S3. This regime is found in the FA$_{0.9}$Cs$_{0.1}$PbI$_{2.8}$Br$_{0.2}$ crystal studied in the present paper. The centrosymmetric cubic structure of this perovskite crystal does not allow for the Dyakonov-Perel mechanism and the anisotropic exchange interaction of localized carriers~\cite{Dzhioev2002}. Therefore, in the following we consider the nuclear spin fluctuations as the only origin of the random fields.

Even time-resolved optical orientation techniques do not always allow one to determine, which of the spin relaxation regimes discussed above is realized in a specific sample. However, this becomes possible by applying external magnetic fields. 
In the regime of short correlation time, the application of a magnetic field $B_\text{V}$ in Voigt geometry, i.e. perpendicular to the wave vector of the exciting light, results in Larmor precession of the electron (hole) spin with the frequency $\omega_\text{L,V} = \mu_\text{B} g B_\text{V}/\hbar$. Here $\mu_\text{B} $ is the Bohr magneton, $g$ is the $g$-factor, and $\hbar$ is the Planck constant. In case of excitation by short pulses of light, the mean spin as function of time demonstrates decaying oscillations:
\begin{equation}
\label{eq_StBV}
\langle S(t) \rangle = \langle S(0) \rangle \cos(\omega_\text{L,V}t) \exp(-t/T_\text{s}).
\end{equation}
For continuous-wave (CW) excitation, this dependence should be time-averaged with the recombination probability:
\begin{equation}
\label{eq_StBV}
P_\text{r}(t) = \frac{1}{\tau}\exp(-t/\tau),
\end{equation}
where $\tau$ is the electron or hole lifetime. This results in a Lorentzian dependence of the mean spin on the magnetic field:
 \begin{equation}
\label{eq_SCW}
\langle S_\text{CW}\rangle =  \frac{T_\text{s}}{\tau} \frac{S_0}{1 + \omega_\text{L,V}^2T_\text{s}^2} = \frac{T_\text{s}}{\tau} \frac{S_0}{1 + (T_\text{s} \mu_\text{B} g B_\text{V}/\hbar )^2}.
\end{equation}
The suppression of spin polarization under CW excitation by a transverse magnetic field, in this particular case described by Eq.~\eqref{eq_SCW}, is known as the Hanle effect. For the data evaluation we use two parameters: the half-width of the Hanle curve $\Delta_\text{H} = \hbar/(\mu_\text{B} gT_\text{s})$ and the amplitude of the Hanle curve $A_\text{H}$. 

A magnetic field $B_\text{F}$ applied in the Faraday geometry, i.e. parallel to the wave vector of light, does not cause regular spin precession. The field is parallel to the mean spin vector of photoexcited electrons (holes) and no effect on the spin polarization is expected, as shown by the red line in Figure~\ref{tc}b.  However, $B_\text{F}$ can slow down the spin relaxation,  making $\tau_\text{s}$, and therefore $T_\text{s}$, dependent on $B_\text{F}$. This results in an increase of the mean spin $\langle S_\text{CW}\rangle = S_0 T_\text{s}(B_\text{F})/\tau$ (Figure~\ref{tc}c). The effect of an increase of the electron (hole) spin polarization is called the polarization recovery (PR) effect.
 
Typically, the PR effect occurs due to the carrier interaction with the nuclear spin fluctuations, when $\omega_\text{L,F} =  \mu_\text{B} g B_\text{F}/\hbar$ becomes comparable to $\tau_\text{c}^{-1}$. The amplitude ($A_{\rm PR}$) and the width ($\Delta_{\rm PR}$) of the PR curve are determined by the value of the correlation time  $\tau_\text{c}$ relative to the spin lifetime $T_\text{s,0}$. Here, $T_{\rm {s,0}} = (\tau_\text{s,0}^{-1} + \tau^{-1})^{-1}$ accounts for the carrier recombination time ($\tau$) and spin relaxation time $\tau_\text{s,0}$, governed by the spin relaxation mechanisms, disregarding the interaction with the nuclear spins. Three cases determined by different values of the ratio $T_\text{s,0}/\tau_\text{c}$ are illustrated schematically in Figures~\ref{tc}d,e,f, based on the theoretical consideration in the Supporting Information, S3. They are considered for the intermediate regime of the correlation time ($\omega_\text{L,r}\tau_\text{c} \gg 1$, $\tau_\text{c} < \tau_\text{c,N}$), which dynamics at zero magnetic field is illustrated in Figure~\ref{dynamics}b. Further, in this paper we consider the correlation time of the nuclear spin fluctuations to be the longest compared to the correlation time of carriers and the carrier spin relaxation time from mechanisms other than the interaction with the nuclear spins.
 

In this regime, the random fields of the nuclear spin fluctuations prevent free Larmor precession of the electron (hole) spin about the external magnetic field. Instead, application of the external field changes the average angle between the electron (hole) spin at the moment of their creation by light and the total field (fluctuation + external) that acts on the spin. In the Faraday geometry, this angle turns closer to zero, which results in an increase of the average spin after its projection on the total field. In the Voigt geometry, the angle turns closer to 90$^\circ$, which results in a reduction of the average spin after projection on the total field. As a result, both the polarization recovery and the Hanle effects occur in about the same range of external magnetic fields. 
 
In the case where the correlation time exceeds $T_\text{s,0}$ ($T_\text{s,0} < \tau_\text{c}$), shown in Figure \ref{tc}d, the widths of the PR and Hanle curves are very close to each other as they are determined by the root mean square field of the nuclear spin fluctuations. The amplitudes of the two effects are related by $A_{\rm PR}/A_{\rm H} = 2/1$. A shortening of the correlation time, as illustrated in Figure \ref{tc}e ($T_\text{s,0} = 5 \tau_\text{c}$) and Figure \ref{tc}f ($T_\text{s,0} = 50 \tau_\text{c}$), is accompanied by an increase of the ratio of amplitudes $A_{\rm PR}/A_{\rm H}$. Also, the PR width increases significantly from 20~mT to 120~mT, while the Hanle width shows the opposite, but much weaker trend decreasing from 20~mT to 16~mT.    
 
In the intermediate regime of correlation time, the spin relaxation time cannot be directly evaluated from the Hanle and PR curves. To do this, one should measure the decay time of the spin polarization under pulsed excitation. In the Faraday geometry, it is approximately equal to the correlation time at zero field and increases with growing $B_\text{F}$, doubling when $B_\text{F}$ reaches the root mean square field of the nuclear spin fluctuations (see Supporting Information, S3 for details).

\begin{figure*}[hbt]
\begin{center}
\includegraphics[width = 18cm]{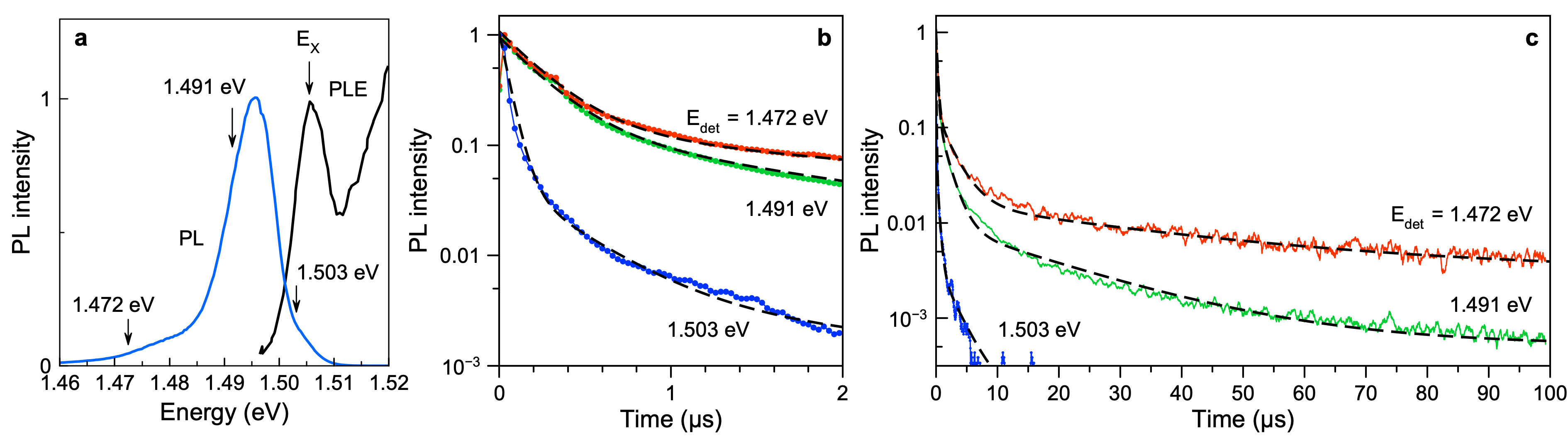}
\caption{\label{fig1} Optical properties of the FA$_{0.9}$Cs$_{0.1}$PbI$_{2.8}$Br$_{0.2}$ crystal at $T = 1.6$~K. (a) Photoluminescence spectrum (blue line) at $T = 1.6$\,K. Black line is photoluminescence excitation spectrum detected at $E_\text{det} = 1.496$~eV. $E_\text{X}$ denotes the exciton resonance.  (b,c) Time-resolved PL measured at various energies in 2~$\mu$s and 100~$\mu$s ranges with $E_\text{exc} = 2.33$\,eV at the laser repetition rate of 10~kHz. The PL dynamics are fitted with a multi-exponential function with the decay times given in Table~\ref{tab:St1}. The fits are shown by the black dashed lines.}
\end{center}
\end{figure*}

\begin{table*}[htb]
\caption{Recombination times in the FA$_{0.9}$Cs$_{0.1}$PbI$_{2.8}$Br$_{0.2}$ crystal measured at different spectral positions for $T = 1.6$\,K.}
\label{tab:St1}
\begin{center}
\begin{tabular*}
{0.55\textwidth}{@{\extracolsep{\fill}} |>{\centering\arraybackslash} m{0.1\textwidth} |>{\centering\arraybackslash} m{0.1\textwidth}|>{\centering\arraybackslash} m{0.1\textwidth}|>{\centering\arraybackslash} m{0.1\textwidth}|>{\centering\arraybackslash} m{0.1\textwidth}|}
\hline
$E_\text{det}$\,(eV) & $\tau_1$\,(ns)& $\tau_2$\,(ns)& $\tau_3$\,($\mu$s)& $\tau_4$\,($\mu$s)\\
\hline
1.503 & 55 & 370 & 3.0 & - \\
\hline
1.491 & -& 250 & 1.6 & 20 \\
\hline
1.472 & - & 200 & 3.3 & 44\\
\hline
\end{tabular*}
\end{center}
\end{table*}

Finally, when studying the effects of optical spin orientation, it is important to take into account the dynamic polarization of nuclear spins (DNP) by the photoexcited electrons and holes. The dynamically polarized nuclear spins create an Overhauser field that affects the PR and Hanle curves. In order to avoid that, the measurements should be performed using high-frequency modulation of the circular polarization of the excitation, which is known to suppress DNP~\cite{OO_book_Ch5,Spin_book_2017_Ch12}. In turn, the DNP can be used to distinguish between the electron and the hole contributions to the observed spin polarization effects. For example, in the perovskite crystal studied in this work, the constants of the hyperfine coupling of electrons and holes with all the nuclear species are positive, while the electron and hole $g$-factors have opposite signs~\cite{kirstein2022am}. In the Faraday geometry, where DNP results in the emergence of a mean nuclear spin parallel to the excitation light beam, the Overhauser field for the electrons is antiparallel to that for the holes. As a consequence, the PR curves for electrons and holes are shifted to positive and negative fields, respectively, and opposite to each other.  

\section{Results}

The studied {FA}$_{0.9}$Cs$_{0.1}$PbI$_{2.8}$Br$_{0.2}$ bulk crystal was grown in solution with the inverse temperature crystallization technique~\cite{nazarenko2017}, see Supporting Information, S1. The $\alpha$- or black phase of {FA}$_{0.9}$Cs$_{0.1}$PbI$_{2.8}$Br$_{0.2}$ has a cubic crystal structure at room temperature. Since the crystal shows negligible anisotropies of the electron and hole $g$-factors at cryogenic temperatures~\cite{kirstein2022nc}, we consider its structure as cubic also in these conditions. The experimental geometry with the light wave vector $\textbf{k}\parallel [001]$ is used in the reported optical studies. 

\begin{figure*}[hbt]
\begin{center}
\includegraphics[width = 14cm]{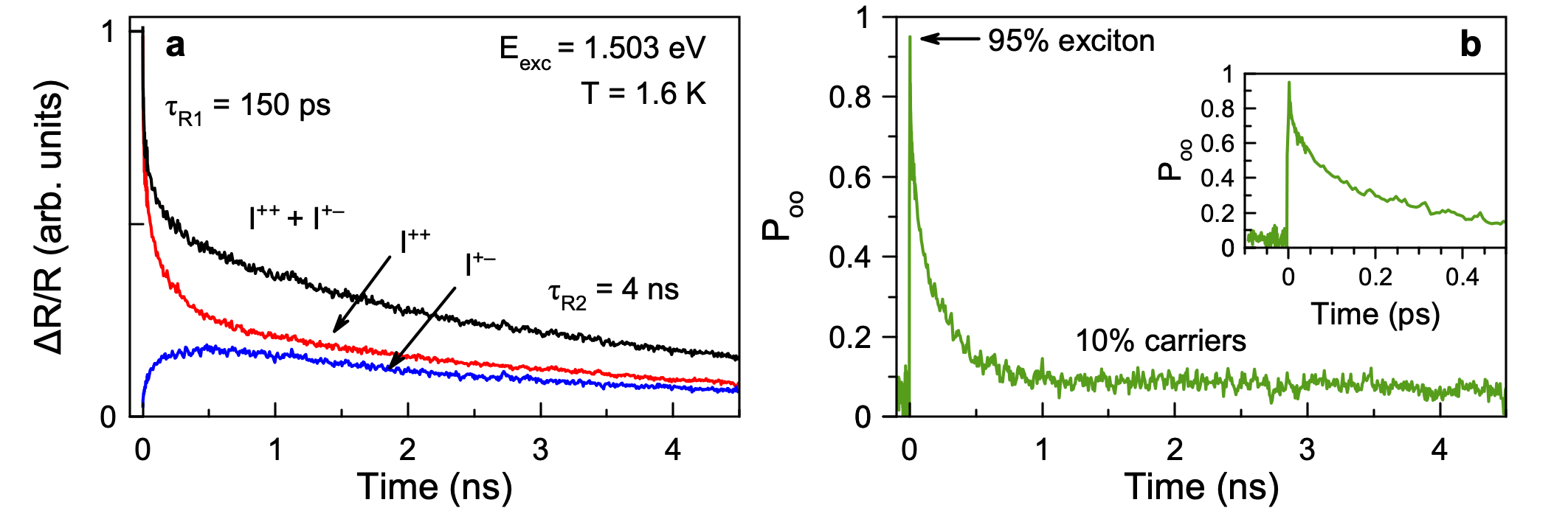}
\caption{\label{dTT}  Dynamics of optical orientation of excitons and carriers at $T = 1.6$~K.  (a) Time-resolved differential reflectivity ($\Delta R/R$) measured at 1.503\,eV for $\sigma^+$ polarized pump excitation. The $\Delta R/R$ dynamics is detected in $\sigma^+$ polarization (red line, $I^{++}$) and $\sigma^-$  polarization (blue line, $I^{+-}$) of the probe. The black line gives $I^{++}+I^{+-}$. The pump power is $45$\,W/cm$^2$ and probe power is $1.8$\,W/cm$^2$.  A two-exponential fit gives the decay times $\tau_\text{R1} = 150$~ps and $\tau_\text{R2} = 4$~ns. (b) Dynamics of the optical orientation degree ($P_\text{oo}$) evaluated from the differential transmission. In insert, the range of short times is zoomed.}
\end{center}
\end{figure*}

\subsection{Optical properties} 

The optical properties of the FA$_{0.9}$Cs$_{0.1}$PbI$_{2.8}$Br$_{0.2}$ crystal reflected by the photoluminescence (PL) are illustrated in Figure~\ref{fig1}a. At the cryogenic temperature of $T = 1.6$~K, the exciton resonance is seen in the photoluminescence excitation (PLE) spectrum at the energy of $E_\text{X} = 1.506$~eV. The exciton binding energy is expected to be very close to the one in FAPbI$_3$ of 14~meV~\cite{galkowski2016}, which allows us to determine the band gap energy in the studied FA$_{0.9}$Cs$_{0.1}$PbI$_{2.8}$Br$_{0.2}$ crystal to be $E_g=1.520$~eV. The photoluminescence spectrum measured under continuous wave excitation shows an emission line with the maximum at 1.495~eV and the full width at half maximum of 8~meV. We showed in previous studies that the recombination of free excitons with the lifetime of $\tau_{\rm X}=60$~ps corresponds to the high energy tail of the PL line, in the vicinity of $E_\text{X}$~\cite{kirstein2022am,XOO}. The PL line is mostly provided by recombination of localized electrons and holes, which are spatially separated in the crystal. This is confirmed by the long recombination dynamics measured at $E_\text{det} = 1.503$~eV and shown in Figures~\ref{fig1}b,c (blue line). The decay comprises three recombination processes with time scales of $\tau_1 = 55$~ns, $\tau_2 = 370$~ns, and $\tau_3 = 3$~$\mu$s, all greatly exceeding the 60~ps time of exciton recombination. We estimate that 50\% of photoexcited carriers recombine at times exceeding 70~ns relative to the moment of their generation. Note that the PL dynamics measured with the time-of-flight technique (see Methods for details) has 30~ns time resolution and, therefore, picosecond exciton dynamics cannot be assessed thereby. The PL dynamics demonstrate dispersion of the recombination times, varying from a nanosecond at energies close to the exciton resonance up to tens of microseconds at lower energies, as shown in Figures~\ref{fig1}b,c. The recombination times for different detection energies are collected in Table~\ref{tab:St1}.  The recombination of electron-hole pairs dominates at the PL maximum and is considerable in the high energy tail, where it overlaps with the exciton emission. The spectral overlap of the exciton and carrier recombination calls for care in interpreting the measured spin-dependent properties. These contributions can be separated in time domain and also by spin phenomena in magnetic field, as we show in this paper.

\subsection{Spin dynamics measured by polarized differential reflectivity}

Time-resolved differential transmission, in which the circularly polarized pump light generates spin-oriented excitons and/or carriers and the circularly polarized probe detects their dynamics, has been applied already for lead halide perovskites: polycrystalline films~\cite{Giovanni2015,zhou2020,Sheng2015}, 2D materials~\cite{Bourelle2020,Giovanni2019,Tao2020}, and nanocrystals~\cite{Rivett2018,Strohmair2020,Meliakov2023}. We use it here as well to study the spin dynamics in the FA$_{0.9}$Cs$_{0.1}$PbI$_{2.8}$Br$_{0.2}$ crystal. Figure~\ref{dTT}a shows the dynamics of $\Delta R/R$ signals measured with a $\sigma^+$ polarized pump and detected with either a $\sigma^+$ (red line, $I^{++}$) or $\sigma^-$ (blue line, $I^{+-}$) polarized probe. In this experiment the laser photon energy is set to 1.503~eV, in the vicinity of the free exciton energy. 
\begin{figure*}[htb]
\begin{center}
\includegraphics[width = 16cm]{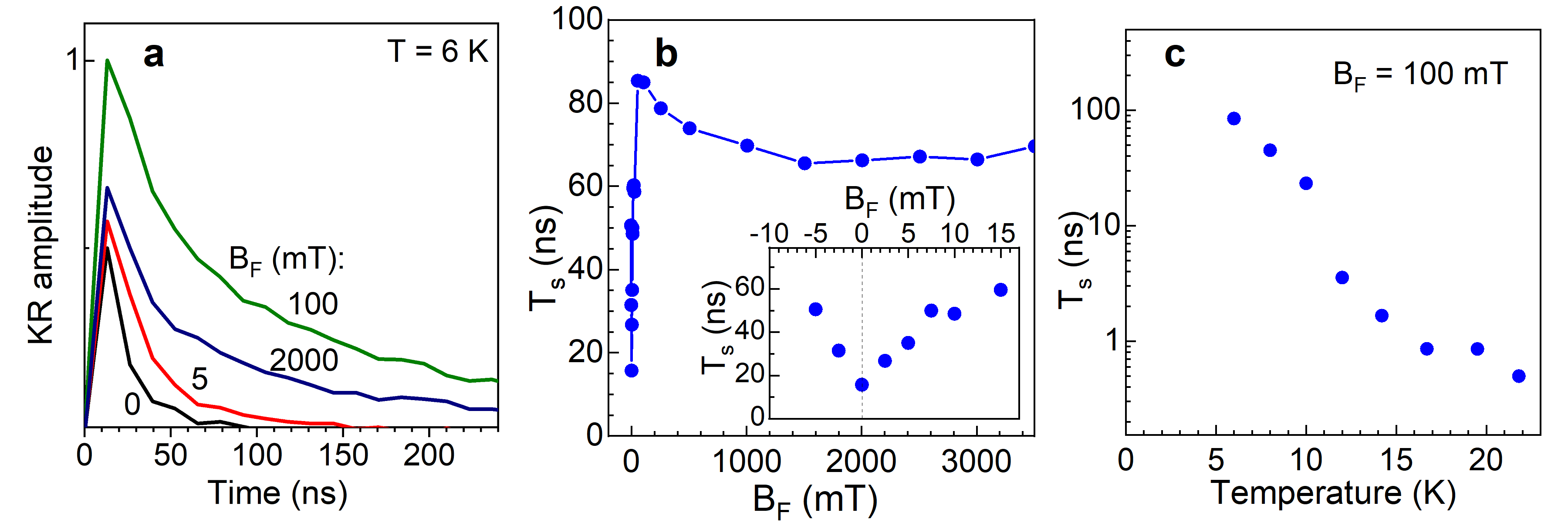}
\caption{\label{fig_T1_B} Spin dynamics of carriers measured by time-resolved Kerr rotation. (a) Dependence on magnetic field applied in the Faraday geometry. $T=6$~K, $E_\text{exc} = E_\text{det} = 1.510$~eV, $P_\text{pump} = 0.04$~W/cm$^2$, and $P_\text{probe} = 2$~W/cm$^2$, pulse repetition period 1050~ns. (b) Longitudinal spin relaxation time $T_{\rm s}$ as function of $B_\text{F}$. Line is a guide for the eye. In the insert the range of weak magnetic fields is zoomed. (c) Temperature dependence of $T_{\rm s}$ measured at $B_{\rm F}=100$~mT. }
\end{center}
\end{figure*}

Let us first discuss the population dynamics corresponding to the applied experimental conditions, namely the used excitation density and $T = 1.6$~K. The population dynamics are shown by the black line in Figure~\ref{dTT}a, which is the sum of the polarized signals $I^{++}+I^{+-}$. The dynamics can be fitted with a bi-exponential decay with times $\tau_\text{R1} = 150$~ps and $\tau_\text{R2} = 4$~ns. We assign the initial short time to the lifetime of the free excitons, which is longer than $\tau_{\rm X}=60$~ps measured under weaker excitation densities. Such increase of the exciton lifetime has been reported in Ref.~\onlinecite{XOO}. The longer time of 4~ns is related to the recombination dynamics of the localized electrons and holes. Note that in this experiment the time range is limited to 4.5~ns, so that we are not able to detect longer recombination times of carriers, which are present in the studied FA$_{0.9}$Cs$_{0.1}$PbI$_{2.8}$Br$_{0.2}$ crystal, see Figure~\ref{fig1}c.

The dynamics of the optical orientation degree, $P_{\rm{oo}}(t)$, shown in Figure~\ref{dTT}b, is evaluated as 
\begin{equation}
\label{eq1}
P_{\rm{oo}}(t) = \frac{I^{++}(t) - I^{+-}(t)}{I^{++}(t) + I^{+-}(t)}.
\end{equation}
$P_\text{oo}$ reaches 0.95 (95\%) right after the pump pulse action, which is very close to the ultimate limit of unity (100\%), achievable in lead halide perovskites due to their band structure, see model analysis in Ref.~\onlinecite{XOO}. The observed value is close to  $P_{\rm{oo}}=0.85$, which we measured for excitons in the same crystal by means of time-resolved PL~\cite{XOO}. The optical orientation degree decays within the exciton lifetime of 150~ps to the value of 0.1 (10\%), which does not decrease further within 4.5~ns, evidencing the long spin relaxation time of localized carriers. This agrees well with results on optical orientation measured by time-resolved PL, where the carrier optical orientation is 0.6 using a lower excitation density of $0.01$~W/cm$^2$ and decreases down to 0.1 for the higher density of $0.6$~W/cm$^2$, see Figure~S3 in Ref.~\onlinecite{XOO}. 

\subsection{Spin relaxation measured by time-resolved Kerr rotation}

Time-resolved Kerr rotation is a powerful tool to study the spin dynamics of electrons and holes in perovskites~\cite{belykh2019,kirstein2022am,kirstein2022mapi}. The carrier spin coherence is photogenerated via the charged exciton (trion) states, so that the spin dynamics of excitons is typically not revealed in these experiments. The extended pump-probe Kerr rotation method allows one to measure spin dynamics exceeding the repetition period of pulsed laser, which is 13.2~ns in our experiments (for details see Methods). We examined with this technique  FA$_{0.9}$Cs$_{0.1}$PbI$_{2.8}$Br$_{0.2}$ crystals with the focus on the coherent spin dynamics in magnetic fields applied perpendicular to the light wave vector (Voigt geometry)~\cite{kirstein2022am}. Here, we use it to study the spin dynamics in longitudinal magnetic fields (Faraday geometry) characterized by the longitudinal spin life time $T_{\rm s}(B_{\rm F})$. 

Kerr rotation dynamics traces measured in various magnetic fields at $T=6$~K are shown in Figure~\ref{fig_T1_B}a. One can see that the shortest decay during 16~ns occurs at zero magnetic field and it slows down with increasing field.  The evaluated $T_{\rm s}(B_{\rm F})$ times from using single-exponential fits are plotted in Figure~\ref{fig_T1_B}b as function of the magnetic field. As mentioned, at zero field $T_{\rm s}(0)=16$~ns, which increases up to 85~ns in $B_{\rm F}=100$~mT and varies only slightly for a further field increase up to 3.5~T. In line with the discussions in the Introduction, such behavior is typical for the carrier spin dynamics in semiconductors, where the carrier hyperfine interaction with nuclear spins is strong. This interaction provides spin relaxation of localized carriers, whose spins precess about the Overhauser field of the nuclear spin fluctuations. This spin relaxation mechanism is suppressed in longitudinal magnetic fields exceeding the Overhauser field.  Note that the recombination time $\tau$ and the carrier spin relaxation time via mechanisms other than the random fields of the nuclear fluctuations $\tau_{\rm s,0}$ are not expected to be modified by such weak magnetic fields. Therefore, we conclude that in the whole range of magnetic fields from 0 to 3.5~T these times are greater than or equal to 85~ns. Having in mind the average recombination time of about 70~ns evaluated from the time-resolved PL data in Figures~\ref{fig1}b,c, we suggest that the spin lifetime $T_{\rm s}$ is mainly given by $\tau$ being shorter than the spin relaxation time $\tau_{\rm s,0} > \tau$.

The temperature dependence of $T_{\rm s}$ time measured at $B_{\rm F}=100$~mT shows a decrease from 85~ns at $T=6$~K down to 0.5~ns at 22~K, Figure~\ref{fig_T1_B}c. The decrease can be related to the delocalization of carriers provided by thermal activation, see also Ref.~\cite{kirstein2022am}. More concretely, this behaviour can be related to the shortening of $\tau_\text{c}$ with increasing temperature, e.g. due to a growing probability of thermal delocalization. 

The time-resolved Kerr rotation signal shows spin dynamics over times greatly exceeding the exciton lifetime of 60~ps. It is evident that the signal originates from the spin dynamics of long-living resident carriers. The electron and hole contributions to the Kerr signal can be separated through the spin relaxation times, if they are strongly different. However, the presented signal is close to mono-exponential, suggesting that electrons and holes have similar spin relaxation times. The contribution of electrons and holes can be distinguished in the time-resolved Kerr rotation signal in transverse magnetic field through their different Larmor precession frequencies, as shown in Ref.~\cite{kirstein2022am} for FA$_{0.9}$Cs$_{0.1}$PbI$_{2.8}$Br$_{0.2}$ crystals. In PL experiments with continuous-wave laser excitation, the electron and hole contributions to the optical orientation degree can be distinguished by means of magnetic fields applied in the Faraday and Voigt geometries, as we show below.
\begin{figure*}[htb]
\begin{center}
\includegraphics[width = 18cm]{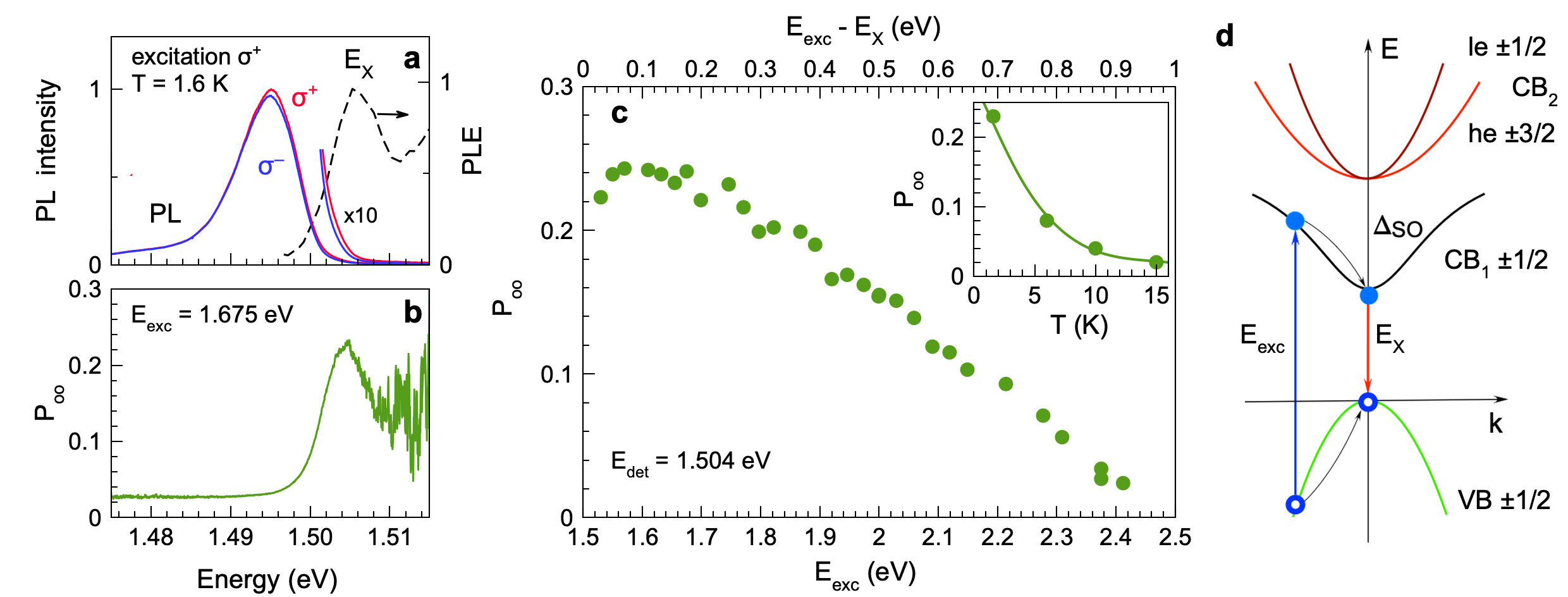}
\caption{\label{OO_intro} Optical orientation in the FA$_{0.9}$Cs$_{0.1}$PbI$_{2.8}$Br$_{0.2}$ crystal under CW excitation at $T = 1.6$~K. (a) Red and blue lines are the photoluminescence spectra measured in $\sigma^+$ and  $\sigma^-$ polarization, respectively, for $\sigma^+$ polarized excitation. $E_\text{exc} = 1.675$~eV  and $P = 0.01$~W/cm$^2$. Dashed line shows the PLE spectrum from Figure~\ref{fig1}a for comparison. (b) Spectral dependence of the optical orientation degree. (c) Dependence of  $P_{\rm oo}$ on the excitation energy detected at $E_\text{det} = 1.504$~eV.  Upper axis shows the detuning from the exciton resonance $E_\text{exc} - E_\text{X}$. The temperature dependence of $P_\text{oo}$ measured at $E_\text{exc} = 1.675$~eV is shown in the insert. (d) Sketch of the band structure of lead halide perovskites with cubic symmetry. VB and CB$_1$ denote the valence and conduction bands with electron and hole spins $1/2$. The CB$_2$ band consisting of the heavy (he) and light (le) electron subbands is shifted from CB$_1$ by the spin-orbit splitting $\Delta_{\rm SO}$. }
\end{center}
\end{figure*}

\subsection{Optical orientation of carriers in photoluminescence under continuous-wave excitation}

The investigation of optical spin orientation of carriers and excitons by means of polarized PL under CW laser excitation is a classical approach in the spin physics of semiconductors~\cite{OO_book,Spin_book_2017}. For this technique the spin signal is time integrated, also the laser photon energy needs to be detuned from the detection energy. In conventional III-V and II-VI semiconductors the detuning has to be rather small, because the carriers loose spin polarization during their energy relaxation. In the lead halide perovskites, due to the different band structure, the conditions and phenomenology of optical orientation are very different~\cite{XOO}. The conduction and valence bands, which form the band gap, are simple and have spin $1/2$, see Figure~\ref{OO_intro}d. This allows one to reach 100\% spin polarization for electrons and holes right after their generation by circularly polarized excitation, even for large detunings exceeding 0.5~eV. Also the spin relaxation during the carrier energy relaxation is inefficient~\cite{Xu2024}. This has been demonstrated by time-resolved PL on FA$_{0.9}$Cs$_{0.1}$PbI$_{2.8}$Br$_{0.2}$ crystals~\cite{XOO}.

As example, we show in Figure~\ref{OO_intro}a the $\sigma^+$ and $\sigma^-$ polarized PL spectra measured under CW excitation with  $\sigma^+$ polarization. The excitation energy is $E_\text{exc} = 1.675$~eV corresponding to the relatively large detuning from the exciton resonance of $E_\text{exc} - E_\text{X} = 0.17$~eV. One can see that the  $\sigma^+$ polarized emission is stronger than the $\sigma^-$ polarized one, evidencing the presence of optical orientation. Its degree, $P_\text{oo}$, is calculated as:
\begin{equation}
\label{eq2}
P_{\rm{oo}} = \frac{I^{++}- I^{+-}}{I^{++} + I^{+-}}.
\end{equation} 

The spectral dependence of $P_\text{oo}$ shown in Figure~\ref{OO_intro}b has a maximum of 0.25 at the energy of 1.504~eV, which is close to the free exciton energy of 1.506~eV in the PLE spectrum. This measured optical orientation is contributed by electrons, holes, and excitons, and the relative ratio of the carrier and exciton contributions changes spectrally. We show below, how these contributions can be distinguished and identified by applying magnetic fields. 

Figure~\ref{OO_intro}c shows the $P_\text{oo}$ detected at 1.504~eV, which corresponds to the maximum of the $P_\text{oo}(E_{\rm det})$ dependence, for a large range of excitation energy detunings from 5~meV up to 900~meV. The value of 0.25 stays constant for detunings up to 0.25~eV, then it gradually decreases to 0.02 at 0.9~eV. This dependence is similar to the one that we measured for the short-living excitons in the same material, while the exciton optical orientation maximum degree reaches 0.85~\cite{XOO}. 

On the basis of these and previous results, the following scenario can be suggested. Electrons and holes photogenerated at large detunings have large $k$-vectors. They do not form excitons, but rather become separated in space due to the opposite momenta of the electron and hole photogenerated by absorption of one photon. Thereafter, fast energy relaxation toward the band edges occurs without spin relaxation due to the suppression of the Dyakonov-Perel mechanism. Electrons and holes at the band edges can form excitons with fast recombination time or get localized separately from each other and recombine during long times. Spin relaxation of localized electrons and holes takes place at the band edges, where, as we show below, their spin relaxation at cryogenic temperatures is provided by the hyperfine interaction with nuclear spin fluctuations. 

The optical orientation degree has a strong temperature dependence as shown in the insert of Figure~\ref{OO_intro}c. It decreases from 0.23 to 0.02 with increasing temperature up to 15~K. We observed a similarly strong temperature dependence in time-resolved Kerr rotation experiments for the spin dephasing times of carriers ($T_2^*$) in various lead halide perovskite crystals~\cite{belykh2019,kirstein2022mapi,kirstein2022am}. We explain this finding by thermal delocalization of electrons and holes that shortens the correlation time $\tau_\text{c}$.
\begin{figure*}[hbt]
\begin{center}
\includegraphics[width = 18cm]{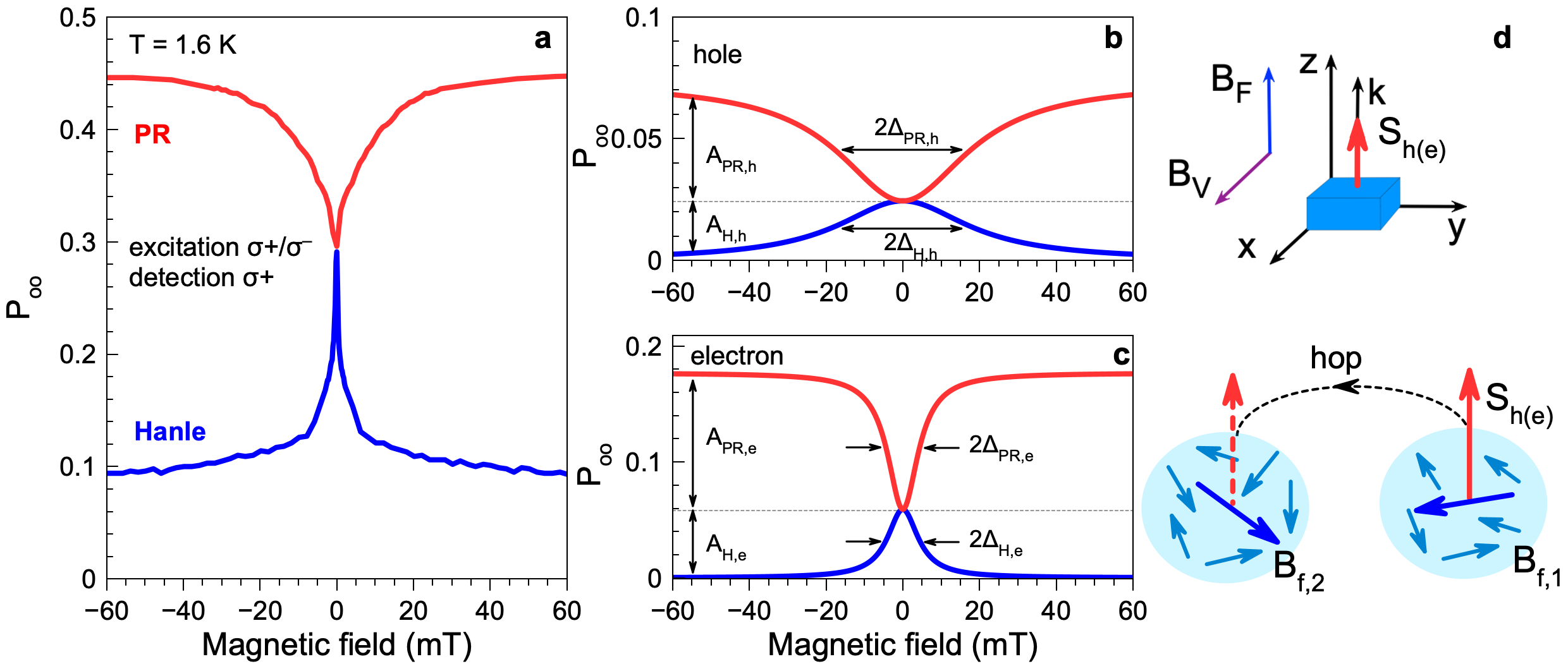}
\caption{\label{Hanle_PRC} Optical orientation of localized carriers in Faraday and Voigt magnetic fields. 
(a) Hanle curve and polarization recovery (PR) curve for helicity-modulated excitation with $f_\text{m} = 50$\,kHz frequency and $\sigma^+$ detection. $T = 1.6$~K, $E_\text{exc} = 1.669$~eV, $E_\text{det} = 1.504$~eV, and $P = 0.5$~W/cm$^2$.  (b) and (c) give the hole and the electron contributions to the Hanle curve (blue) and the PR curve (red). 
(d) Schematics of experimental geometries. The spins of the localized carriers are oriented along the light $k$-vector. Faraday geometry  ($\textbf{B}_\text{F} \parallel \textbf{k}$) is implemented for the PR measurement, and Voigt geometry ($\textbf{B}_\text{V} \perp \textbf{k}$) is for the Hanle experiment. 
(c) Schematics of static nuclear spin fluctuations acting on carrier hopping between localization sites. $B_\text{f,1(2)}$ is the effective field of the nuclear fluctuations on the localization sites.}
\end{center}
\end{figure*}

\subsection{Hanle and polarization recovery effects in magnetic field}

In order to separate the electron, hole, and exciton contributions to the measured optical orientation degree and to gain deeper insight into spin relaxation mechanisms, we measure the effects of an external magnetic field on $P_\text{oo}$. $P_\text{oo}$ decreases in magnetic field applied perpendicular to the light $k$-vector ($\mathbf{B}_\text{V} \perp \mathbf{k}$, Voigt geometry), which is known as the Hanle effect~\cite{OO_book}, while in longitudinal magnetic field ($\mathbf{B}_\text{F} \parallel \mathbf{k}$, Faraday geometry) it increases, which is known as polarization recovery (PR) effect~\cite{smirnov2020}. 

The Hanle curve detected at $E_{\rm det}=1.504$~eV is shown in Figure~\ref{Hanle_PRC}a by the blue trace. It is important to note, that in this experiment, in order to avoid effects related to dynamic nuclear polarization, we modulate the helicity of the excitation laser at the frequency of 50~kHz, and detect the signal in $\sigma^+$ polarization, see Methods. A detailed analysis shows that the Hanle curve has four contributions. The very broad (nearly constant in this range of magnetic fields) background with an amplitude of 0.10 is due to the optical orientation of excitons having the short lifetime of 60~ps. For this value the estimated the half width at half maximum (HWHM) of the Hanle curve for the exciton is $\Delta_\text{H,X} = 160$~mT.

\begin{table*}[hbt]
\caption{Parameters of the Hanle and PR effects in the FA$_{0.9}$Cs$_{0.1}$PbI$_{2.8}$Br$_{0.2}$ crystal for $T = 1.6$~K, measured at $E_\text{det} =1.504$~eV and $1.497$~eV. $^*$ indicates estimation for the exciton recombination and spin relaxation times.  }
\label{tab:t1}
\begin{center}
\begin{tabular*}
{1\textwidth}{@{\extracolsep{\fill}}|>{\centering\arraybackslash} m{0.1\textwidth}|>{\centering\arraybackslash} m{0.1\textwidth} |>{\centering\arraybackslash} m{0.1\textwidth}|>{\centering\arraybackslash} m{0.12\textwidth}|>{\centering\arraybackslash} m{0.12\textwidth}|>{\centering\arraybackslash} m{0.12\textwidth}|>{\centering\arraybackslash} m{0.12\textwidth}|}
\hline
  \multicolumn{1}{|c|}{\multirow{1}{*}{ }}& \multicolumn{6}{c|}{\multirow{1}{*}{$E_\text{det} =1.504$~eV}}\\
\hline
& $A_\text{H}$ &$A_\text{PR}$ & $P_\text{max}$ & $ \Delta_\text{H}$~(mT) &  $ \Delta_\text{PR}$~(mT) &  $ \Delta_\text{N}$~(mT)  \\
\hline
Electron & 0.06 & 0.12&  0.18 & 5 & 5 & 2.5 \\
\hline
Hole & 0.02& 0.05& 0.07 & 19 & 19 & 9.5 \\
\hline
Exciton & 0.10 &-& - & 160$^*$ & - & -\\
\hline
  \multicolumn{1}{|c|}{\multirow{1}{*}{ }}& \multicolumn{6}{c|}{\multirow{1}{*}{$E_\text{det} =1.497$~eV}} \\
\hline
Electron &0.007& 0.014& 0.02 & 4 & 4 & 2.0 \\
\hline
Hole &0.005 & 0.010& 0.015 & 19 & 19 & 9.5 \\
\hline
Exciton & 0 & 0 & - & - & - & -\\

\hline
\end{tabular*}
\end{center}
\end{table*}

The exciton background disappears when the detection energy is detuned from the exciton resonance to 1.497~eV, where only narrow signals from localized carriers remain (see Supporting Information, Figure~S1a). We fit the Hanle curve on top of the exciton background with a sum of three Lorentz contours and evaluate their amplitudes and widths with Eq.~\eqref{eq_SCW}, see the parameters in Table~\ref{tab:t1}. One can sum up the different contributions, if their amplitudes are small, see Supporting Information, S3 and Eq. (S16) for details. We assign two of them to localized electrons ($A_\text{H,e} = 0.06$, $\Delta_\text{H,e} = 5$\,mT) and holes ($A_\text{H,h} = 0.02$, $\Delta_\text{H,h}  = 19$\,mT), respectively. They are shown individually by the blue lines in Figure~\ref{Hanle_PRC}b for holes and in Figure~\ref{Hanle_PRC}c for electrons. The assignment is based on the considerably stronger hyperfine interaction of holes, compared with the one of electrons, which is a specific of the lead halide perovskites~\cite{kirstein2022am}. This results in broader Hanle and PR curves, and also in a difference in the dynamic nuclear polarization, as we show in Sec.~IIf.  The third contribution to the Hanle curve has a very small width of $0.4$~mT and an amplitude of 0.1. 


The PR effect measured in the Faraday geometry at $E_{\rm det}=1.504$~eV is shown in Figure~\ref{Hanle_PRC}a by the red line. It has three components and can be fitted by:
\begin{eqnarray}
\label{PRC_class}
P_\text{oo}(B_\text{F}) =  A_\text{max} - A_\text{PR} \frac{\Delta^2_\text{PR}}{B_\text{F}^2 + \Delta^2_\text{PR}}.
\end{eqnarray}
Here, $A_\text{max}$ is $P_\text{oo}(B_\text{F} \rightarrow \infty)$ and $\Delta_{\text{PR}}$ is the HWHM of the PR curve.  We evaluate  $A_{\text{PR},\text{h}} = 0.12$, $\Delta_{\text{PR,e}}  = 5$\,mT for the localized electrons and  $A_{\text{PR},\text{e}} = 0.05$, $\Delta_{\text{PR,h}}  = 19$\,mT for the localized holes. The individual PR curves are shown by the red lines in Figure~\ref{Hanle_PRC}b for holes and in Figure~\ref{Hanle_PRC}c for electrons. Note, that similar to the Hanle curve, the PR curve shows a very narrow contour with the width of $0.4$~mT and amplitude of 0.01, that we assign to weakly localized electrons.
\begin{figure*}[htb]
\begin{center}
\includegraphics[width = 18cm]{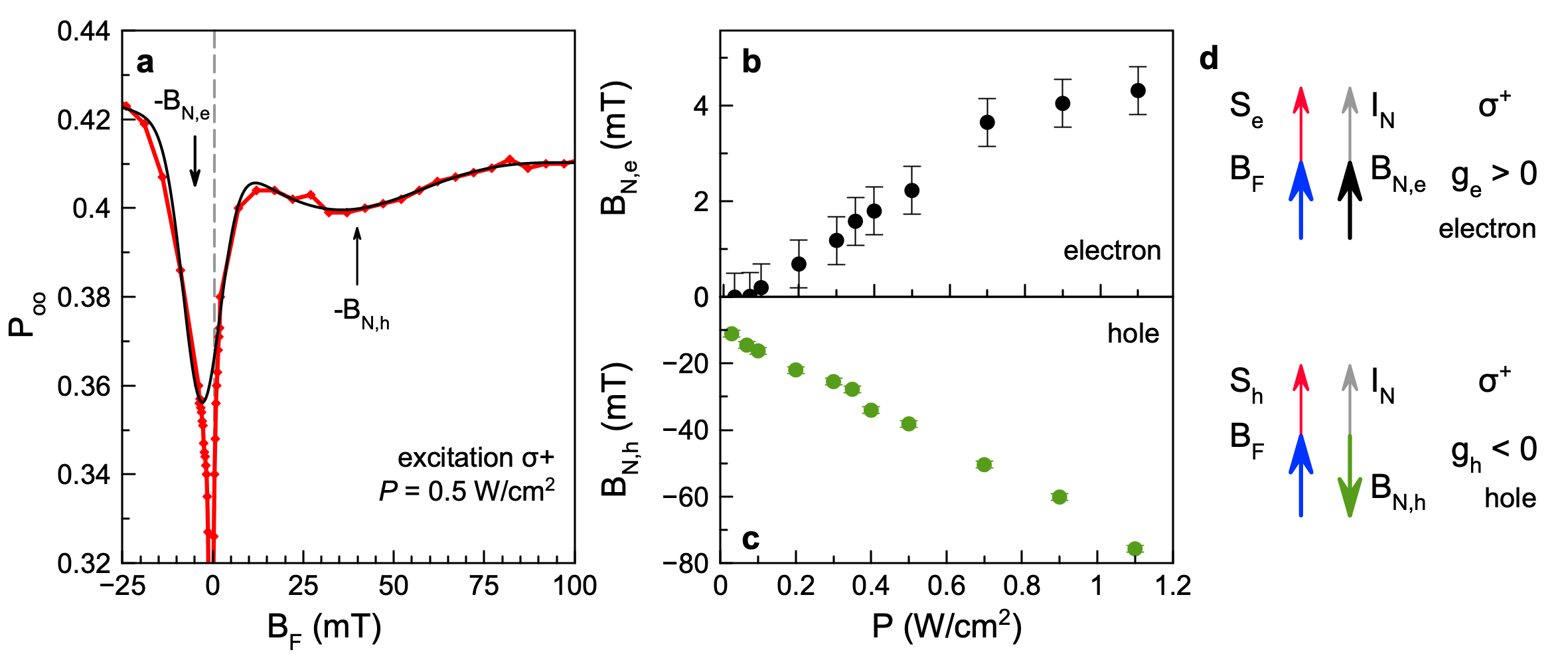}
\caption{\label{DNP} Dynamic nuclear spin polarization in the FA$_{0.9}$Cs$_{0.1}$PbI$_{2.8}$Br$_{0.2}$ crystal at $T = 1.6$~K. (a) PR (red symbols and line) measured at the excitation intensity of $P = 0.5$~W/cm$^2$ for $\sigma^+$ excitation for $E_\text{exc} = 1.675$~eV, $E_\text{det} = 1.504$~eV.  Black line shows the fit with two Lorentzians having minima at $B_\text{N,e} = 2.2$\,mT for the electrons and at $B_\text{N,h} = -38.2$\,mT for the holes. (b,c) Dependences of the Overhauser fields for electrons and holes on the excitation density. (d) Schematics of the $B_\text{N}$ orientation for $\sigma^+$ polarized excitation in lead halide perovskite semiconductor with $g_\text{e} > 0$ and $g_\text{h} < 0$.}
\end{center}
\end{figure*}

The amplitudes and widths of the Hanle and PR curves evaluated by Eqs.~\eqref{eq_SCW},\eqref{PRC_class} are collected in Table~\ref{tab:t1}.  One can see that both for electrons and holes the HWHM of the Hanle and PR curves are equal, indicating that the widths are determined by the interaction with the nuclear spin fluctuations in the limit of an intermediate correlation time in the electron-nuclear spin system~\cite{smirnov2020,Korenev} ($\omega_\text{L,r}\tau_\text{c} \gg 1$, $\tau_\text{c}<\tau_\text{c,N}$), and for the case of $T_\text{s,0} < \tau_\text{c}$, which is schematically illustrated in Figure~\ref{tc}d.   

For comparison, we give in Table~\ref{tab:t1} also the parameters measured at 1.497~eV, which is  close to the PL maximum. The experimental data for this detection energy are shown in the Supporting Information, Figure~S1. One can see that the characteristic widths at both detection energies are about the same, meaning that the number of nuclear spins interacting with electrons and holes does not change considerably. But the polarization amplitude is smaller at lower energy (close to the PL maximum). The recombination time of the electron-hole pairs is increasing for lower energies providing more time for spin relaxation, leading to a decrease of $P_\text{max}$. 

\subsection{Dynamic nuclear polarization}

Spin oriented electrons and holes can transfer their polarization to the nuclear spin system, which becomes polarized as a consequence. The effect is known as dynamic nuclear polarization~\cite{OO_book,Spin_book_2017}. In turn, the polarized nuclei act back on the carriers via the Overhauser field ($\bf{B}_\text{N}$), which exceeds the characteristic fluctuation field $\Delta_\text{N}$. The DNP effect is expected to be strong in materials with intermediate and long correlation time of carriers. Experimentally, the Overhauser field can be evaluated, e.g., from the shift of the PR curve minimum in experiments under excitation with constant helicity, see Methods. The minimum shifts to the magnetic field value, where the external magnetic field compensates the Overhauser field: ${B_\text{F}} + {B}_\text{N} = 0$.

\begin{figure*}[t!]
\begin{center}
\includegraphics[width = 15cm]{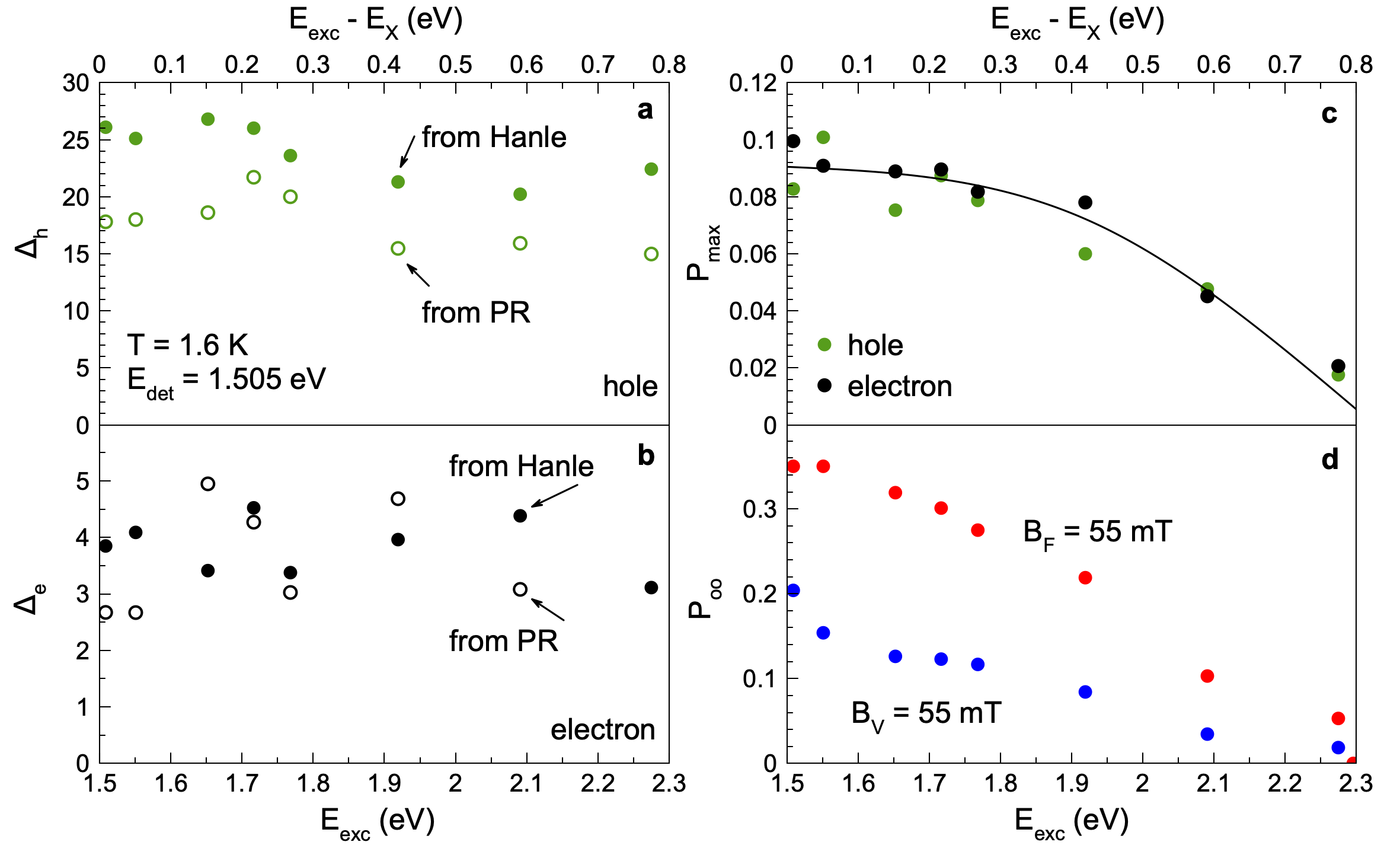}
\caption{\label{detuning} Dependences of optical orientation parameters on excitation energy measured at  $E_\text{det} = 1.505$~eV and $P = 0.01$~W/cm$^2$. (a,b)  $\Delta_\text{PR,h(e)}$ and $\Delta_\text{H,h(e)}$  evaluated from PR (open symbols) and Hanle (closed symbols) curves.  (c) $P_\text{max}$ of electron (black circles) and hole (green circles).  Line is guide to the eye.   (d) $P_\text{oo}$ measured at $B_\text{F} =55$~mT (red circles) and $B_\text{V} =55$~mT (blue circles). } 
\end{center}
\end{figure*}

The results of such experiment, measured for the excitation density of $P = 0.5$~W/cm$^2$, are shown in Figure~\ref{DNP}a. By fitting them with two Lorentzians we get obtain for the hole holes $B_\text{N,h} = -38.2$~mT and for the electrons $B_\text{N,e} = 2.2$~mT. With increasing pump power, more spin-oriented carriers are generated and higher spin polarization is transferred to the nuclear spin system. Figures~\ref{DNP}b and \ref{DNP}c show how the Overhauser field for electrons and holes, respectively, increases with growing excitation density from 0.03 up to 1.1~W/cm$^2$. $B_\text{N}$ grows linearly  and reaches $B_\text{N,e} = 4$\,mT and $B_\text{N,h} = -76$\,mT at $P = 1.1$\,W/cm$^2$. This is in agreement with the results on the dynamic nuclear polarization that we measured by the shift of the Larmor precession frequency in time-resolved Kerr rotation experiments for the same material~\cite{kirstein2022am}. 

The opposite signs of $B_\text{N,e}$ and $B_\text{N,h}$, evident in Figure~\ref{DNP}a, are related to the opposite signs of the electron and hole $g$-factors in FA$_{0.9}$Cs$_{0.1}$PbI$_{2.8}$Br$_{0.2}$ crystals~\cite{kirstein2022nc}. The Land\'e $g$-factors of carriers in this material are about isotropic and are in the range $g_{\rm e}=+3.48$ to $+3.60$ and  $g_{\rm h}=-1.15$ to $-1.22$. Figure~\ref{DNP}d illustrates schematically the process of  dynamic nuclear polarization for the electrons and holes. The spin polarization of the electron ($\bf{S}_\text{e}$) and hole ($\bf{S}_\text{h}$) is directed along the $k$-vector of light, according to the selection rules for $\sigma^+$ excitation. The nuclear spin polarization $\langle\bf{I}_\text{N}\rangle$ is co-directed with $\bf{S}_\text{e}$ and $\bf{S}_\text{h}$. The Overhauser field can be expressed as ${\bf{B}}_\text{N,e(h)} = A_\text{hf}^\text{e(h)}\langle {\bf{I}}_\text{N} \rangle/\mu_\text{B}g_\text{e(h)}$. Here $A_\text{hf}^\text{e(h)}>0$ are the hyperfine constants for electrons (holes). As they are positive for both types of carriers~\cite{kirstein2022am}, the direction of the Overhauser field is determined by the sign of carrier $g$-factors.  

\subsection{Carriers energy and spin relaxation}

\begin{figure*}[htb]
\begin{center}
\includegraphics[width = 18cm]{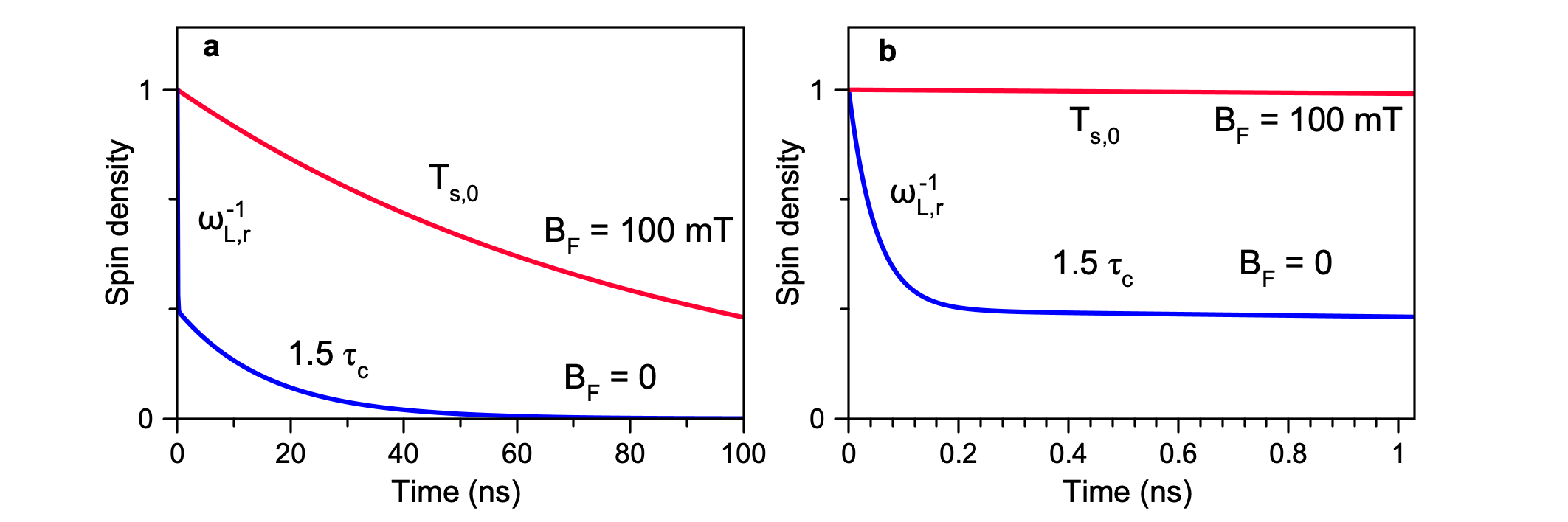}
\caption{\label{diss_tc} Modeled spin density dynamics. (a) Dynamics with the spin lifetime governed by the nuclear spin fluctuations at short time scale and the correlation time at ns-time scales (blue line), corresponding to the Kerr rotation dynamics at $B_\text{F} = 0$, shown in Figure~\ref{fig_T1_B}a. The case of the longitudinal magnetic field exceeding the nuclear spin fluctuations at $B_\text{F} = 100$~mT (red line) is experimentally observed in Figure~\ref{fig_T1_B}a. (b) Zoom of the dependences from panel (a) in the 1~ns range. Model parameters are $\tau_\text{c} = 11$~ns, $T_{\rm s,0}=85$~ns, $\Delta_\text{N,h} = 10$~mT and $g_\text{h} = -1.2$.}
\end{center}
\end{figure*}

We measure the Hanle and PR curves for various detunings of the excitation energy from the exciton resonance in the range of $E_{\rm exc}-E_{\rm X}$ from 5~meV to 0.78~eV and present their parameters in Figure~\ref{detuning}. One can see that the widths $\Delta_\text{PR,h(e)}$ and $\Delta_\text{H,h(e)}$ are nearly independent of the detuning for both electrons and holes. This confirms that the carrier spin relaxation mostly occurs via their interaction with the nuclear spin fluctuations, when the carriers are already localized at the band edges. The dependence of  $P_\text{max}$ on the detuning is shown in Figure~\ref{detuning}c and is very similar for the electrons and holes. Namely, the polarization is constant up to a detuning of 0.25~eV and then slowly decreases to 0.02 at 0.78~eV.  The plateau in the polarization shows that spin relaxation does not occur during carrier  energy relaxation.  The gradual decrease at large detunings for the electrons and holes is due to the deviation of the selection rules from chiral ones far above the R point and a possible depolarization by phonons during the energy relaxation, see Ref.~\onlinecite{XOO} for a detailed analysis. 

For completeness, we present the dependences of the optical orientation on excitation detuning measured in magnetic fields of 55~mT for both field orientations. In the Faraday geometry (red circles in Figure~\ref{detuning}d) this corresponds to the saturation level of the PR signal, which is contributed by electrons, holes, and excitons. In the Voigt geometry (blue circles) the electron and hole contributions are already suppressed and only the exciton optical orientation is measured. The qualitative behavior of both data sets is similar to the dependence of optical orientation at zero magnetic field as shown in Figure~\ref{detuning}c.

\section{Discussion}

We have studied the spin properties including dynamics of electrons and holes in an FA$_{0.9}$Cs$_{0.1}$PbI$_{2.8}$Br$_{0.2}$ crystal, which can be considered as a representative material for the lead halide perovskite semiconductors. For that, optical orientation techniques with time resolution and time averaging in excitation and detection have been used. In the experiments, electrons and holes are photonegerated with large energy detuning from the band gap and relax toward the band edges without loosing their spin orientation. After energy relaxation, the carriers can be bound to excitons or become localized in spatially separated states. In zero magnetic field, all contributions are present in the optical orientation signal with time resolution and in CW experiments. 

In order to identify the spin signals of electrons, holes, and excitons, we measure the Hanle and polarization recovery effects in external magnetic fields. The detailed theoretical consideration of the carrier spin dynamics in the intermediate regime of the correlation time, as presented in the Supporting Information, S3 provides the following equations for the PR and Hanle effects:  

\begin{eqnarray}
\label{PRC_NF}
P_\text{oo}(B_\text{F}) = \frac{(3B^2_\text{F} + 4\Delta_\text{N}^2)P_\text{max}}{3B^2_\text{F} + 4\Delta_\text{N}^2(3 + 2T_{\rm {s,0}}/\tau_\text{c})},
\end{eqnarray}
\begin{eqnarray}
\label{Hanle_NF}
P_\text{oo}(B_\text{V}) = \frac{4\Delta_\text{N}^2P_\text{max}}{3B^2_\text{V}(1 + T_{\rm {s,0}}/\tau_\text{c}) + 4\Delta_\text{N}^2(3 + 2T_{\rm {s,0}}/\tau_\text{c})}.
\end{eqnarray}
Here $\Delta_\text{N}$ is the HWHM of the nuclear fluctuations described by a Gaussian distribution, while $T_{\rm {s,0}} = (\tau_\text{s,0}^{-1} + \tau^{-1})^{-1}$ accounts for carrier recombination and spin relaxation mechanisms other than the  interaction with the nuclear spins. The HWHMs of the PR and Hanle curves in the regime of the intermediate correlation time ($\omega_\text{L,r}\tau_\text{c} \gg 1$ and $\tau_\text{c} < \tau_\text{c,N}$) are:
\begin{equation}
\label{PR_width}
\Delta_\text{PR}^2=4\Delta_\text{N}^2\frac{2T_{\rm {s,0}}+3\tau_c}{3\tau_c},
\end{equation}
\begin{equation}
\label{H_width}
\Delta_\text{H}^2=4\Delta_\text{N}^2\frac{2T_{\rm {s,0}}+3\tau_c}{3T_{\rm {s,0}} + 3\tau_c}.
\end{equation}
Note that these expressions can also be also obtained from Eqs.~(18,19) of Ref.~\onlinecite{smirnov2023si} for long correlation time. 

One can see in Figure~\ref{Hanle_PRC} and from Table~\ref{tab:t1} that the PR and  Hanle curves of the electrons have the same width, also their amplitudes ratio $A_{\rm PR}/A_{\rm H} = 2/1$. The same is valid for the holes. These interdependencies are characteristic for the spin dynamics with intermediate correlations times for $T_{\rm {s,0}} < \tau_\text{c}$ as illustrated in Figure~\ref{tc}d.  In this case, as one can see from Eqs.~\eqref{PR_width},\eqref{H_width}, the nuclear fluctuations can be evaluated as $\Delta_\text{N} = \Delta_\text{PR}/2 = \Delta_\text{H}/2$. This allows us to evaluate for the electrons $\Delta_\text{N,e} = 2.5$~mT and for the holes $\Delta_\text{N,h} = 9.5$~mT.  


Let us analyze the spin dynamics measured by the extended pump-probe Kerr rotation shown in Figure~\ref{fig_T1_B}. The detected signal here is proportional to the spin density, which is induced by the circularly polarized pump. The signal decay with time $T_{\rm s}$ can be contributed by both the carrier spin relaxation and the carrier recombination. The strong increase of $T_{\rm s}$ from 16~ns to 85~ns in a weak magnetic field of about 10~mT, which is comparable  with the characteristic nuclei fluctuations  $\Delta_\text{N}$ measured for this sample, lets us conclude that the random fields of the nuclear fluctuations determine the spin dynamics. In this experiment, performed at $T=6$~K, the condition $T_{\rm {s,0}} > \tau_\text{c}$ is valid, which can be explained by shortening of the correlation time with increasing temperature, compared to the conditions at $T=1.6$~K at which the PR and Hanle effect have been measured under CW excitation.  

A theoretical analysis of the carrier spin dynamics is given in the Supporting Information, S3. The representative modeling based on the experimentally relevant parameters is presented in Figure~\ref{diss_tc}. For $T_{\rm {s,0}} > \tau_\text{c}$, in zero and weak magnetic fields comparable with the  nuclear spin fluctuation ($B_\text{F} \le \Delta_\text{N}$), the spin density dynamics shows a fast decay with decreasing signal amplitude from 1 down to 1/3 on a 100~ps time scale, see the blue line in Figures~\ref{diss_tc}a,b calculated for $B_\text{F}=0$.  Note that this fast decay is not seen in Figure~\ref{fig_T1_B}a due to the low time resolution of the experimental protocol aimed for measuring the long-lasting dynamics. The further decay is slow and is controlled by the correlation time. It is described by Eq.~(S10) of the Supporting Information, which for $B=0$ gives the slow decay time of $1.5\tau_{\rm c}$. This allows us to estimate  $\tau_\text{c} = 11$~ns from the experimentally measured time  $T_{\rm s}(B_\text{F}=0)=16$~ns.

In the limit of $B_\text{F} \gg \Delta_\text{N}$, the spin density is governed by the time $T_{\text{s},0}$, which is contributed by the intrinsic spin relaxation or recombination mechanisms, see the red line in Figures~\ref{diss_tc}a,b. In experiment, the dependence of $T_{\text{s}}$ on magnetic field saturates at about 70~ns and becomes independent of the magnetic field, see Figure~\ref{fig_T1_B}b. This time is close to the average recombination time of about 70~ns evaluated from the time-resolved PL data in Figures~\ref{fig1}b,c. Therefore, we conclude that the spin lifetime $T_{\rm s,0}$ is mainly determined by $\tau$ and that $\tau_{\rm s,0} > \tau$.  The surprisingly long carrier spin relaxation time exceeding 100~ns is due to the suppression of the spin-orbit relaxation mechanisms provided by the unbroken inversion symmetry in cubic perovskites.

\section{Conclusion}

We use both the PR and Hanle effects for addressing the optical orientation in  FA$_{0.9}$Cs$_{0.1}$PbI$_{2.8}$Br$_{0.2}$ crystals and show that only the joint analysis of both effects provides a conclusive interpretation of the experimental data. We develop a theoretical model of carriers hopping and spin relaxation mediated by the hyperfine interaction (Supporting Information, S3), which successfully describes the ratio of Hanle and PR amplitudes and their widths for different ratios of the spin lifetime to the correlation time. In Ref.~\cite{smirnov2020}, a detailed theoretical consideration of other possible mechanisms of carrier-nuclear spin interactions are provided and partly illustrated by experimental results. 

In conventional semiconductors and their nanostructures, the role of localized carrier spins and polarized nuclear spins interacting with the carriers are widely examined in light of spintronics and quantum information applications.  However, for the perovskite semiconductors only first steps have been made along this direction. Recently, the suppression of the nuclear spin fluctuations via creation of a squeezed dark nuclear spin state was demonstrated~\cite{smirnov2023}. The effect is governed by quantum correlations and entanglement between the nuclear spins, prepared optically in FAPbBr$_3$ perovskite crystals. The presented results show that the spin properties of lead halide perovskite semiconductors are competitive with that of conventional semiconductors, and also offer regimes, which are hardly achievable in conventional materials.

\section*{Methods}

\textbf{Samples.}   The studied {FA}$_{0.9}$Cs$_{0.1}$PbI$_{2.8}$Br$_{0.2}$ bulk single crystal was grown out of a solution with the inverse temperature crystallization technique~\cite{nazarenko2017}. Therefore, a solution of CsI, FAI, PbI$_{2}$ and PbBr$_{2}$ is mixed with GBL $\gamma$-butyrolactone solvent. The solution is filtered and heated to $130^\circ$C, so that crystals are formed in the $\alpha$-phase. Single crystals are separated by filtration and drying. The $\alpha$-phase or black phase of {FA}$_{0.9}$Cs$_{0.1}$PbI$_{2.8}$Br$_{0.2}$ has a cubic crystal structure at room temperature. Further growth details are given in the Supporting Information, S1. Since in {FA}$_{0.9}$Cs$_{0.1}$PbI$_{2.8}$Br$_{0.2}$ crystals about isotropic $g$-factors were measured for the electrons and the holes at low temperatures~\cite{kirstein2022nc}, we consider its crystal structure as cubic also under these conditions. The geometry with the light wave vector $\textbf{k}\parallel [001]$ was used in all optical experiments.

\textbf{Magneto-optical measurements.}  For low-temperature optical measurements we use a liquid helium cryostat with the sample chamber temperature variable from 1.6\,K up to 300\,K. At $T=1.6$\,K the sample is placed in superfluid helium, while at $4.2-15$~K it is held in cold helium gas. Magnetic fields up to 200~mT are generated by an electromagnet with one pair of split coils. The electromagnet is rotated by $90^\circ$ to change the experimental geometry between the Faraday geometry ($\textbf{B}_{\rm F} \parallel \textbf{k}$) and the Voigt geometry ($ \textbf{B}_{\rm V} \perp \textbf{k}$). 

\textbf{Time-resolved photoluminescence.} The photoluminescence is excited by a pulsed laser with photon energy of 2.33~eV (532~nm wavelength), pulse duration of 5~ns, repetition rate of 10~kHz, and average excitation power of 8~$\mu$W. The detection energy ($E_\text{det}$) is selected by a 1~m double monochromator (Yobin-Yvon U1000) equipped with a Peltier-cooled GaAs photomultiplier. The signal is detected using an avalanche photodiode and a time-of-flight PC board with a time resolution of 30~ns. 

\textbf{Optical orientation.} Here, the photoluminescence is excited by a continuous-wave laser tunable in spectral range of $1.18-1.85$~eV (Ti:Sapphire laser) or $1.85-2.95$~eV (MixTrain unit providing the sum frequency of the Ti:Sapphire laser and a fiber laser emitting at 1950~nm). The PL spectra are dispersed by a 1~m double monochromator (Jobin-Yvon U1000) equipped with a Peltier-cooled GaAs photomultiplier. The signal intensity is recorded by a FastCom time-of-flight card synchronized with a photoelastic modulator used for switching the circular polarization between $\sigma^+$ and $\sigma^-$ at a frequency of 50~kHz. Depending on the experimental requirements the modulator is used either for the excitation or for detection.  
 
For the modulator placed in the detection optical path, the optical orientation degree is calculated using Eq.~\eqref{eq2}. For the experiments, where the dynamic nuclear polarization should be excluded, we either use low excitation densities or modulate the laser helicity and detect the PL in s fixed polarization, e.g., $\sigma^+$ polarization. The resulting optical orientation degree is calculated as
\begin{equation}
\label{eq_OO_no_DNP}
P_{\rm{oo}} = \frac{I^{++} - I^{-+}}{I^{++} + I^{-+}}.
\end{equation}
Here $I^{++}$ and $I^{-+}$ are the PL intensities for $\sigma^+$ and $\sigma^-$ excitation, respectively, detected in $\sigma^+$ polarization. In the experiments on dynamical polarization of the nuclear spin system, the laser polarization is kept constant ($\sigma^+$) and the modulator is placed in the detection path. 

\textbf{Photoluminescence excitation (PLE).}  In order to determine the energy of the free exciton, we measure a PLE spectrum (corresponding rather closely to an absorption spectrum), where the PL intensity is detected at the PL maximum ($E_{\rm{det}} = 1.496$\,eV) and the continuous-wave Ti:Sapphire laser is tuned in spectral range of $1.497-1.520$~eV.
 
\textbf{Time-resolved differential reflectivity.} A pulsed laser (pulse duration of 1.5~ps, repetition rate of 76~MHz, repetition period of 13.2~ns) is used for this pump-probe experiment. The pump and probe beams have the same photon energy set close to the exciton resonance at 1.503~eV. The pump is $\sigma^+$ circularly polarized, its intensity is modulated by an electro-optical modulator at the frequency of 100~kHz. The probe beam is either $\sigma^+$ or $\sigma^-$  polarized. The probe beam reflected from the sample and the reference laser beam are sent to the balanced photodetector connected with a lock-in amplifier. The differential reflectivity ($\Delta R/R$) dynamics are measured by varying the pump-probe time delay by a mechanical delay line. The temporal resolution is about 2~ps. 

\textbf{Extended pump-probe Kerr rotation.} 
We use the extended pump-probe Kerr rotation technique~\cite{belykh2016} to study the longitudinal spin dynamics of the carriers (Figure~\ref{fig_T1_B}). In these experiments a Ti:Sapphire laser emits a train of 2~ps pulses with a repetition rate of 76~MHz (repetition period $T_\text{R}=13.2$~ns). The laser output is split into the pump and probe beams. In the pump path an electro-optical modulator (EOM) is installed to select single pulses separated by a period of $80 T_\text{R}$. An acousto-optical light modulator (AOM) in the probe path is used to  select single pulses also with the repetition period of $80 T_\text{R}$ at the required delay after the pump pulse. Electronic variation of the delay between the synchronized AOM and EOM (also synchronized with the laser) allows for a change of the delay between the pump and probe pulses in steps of $T_\text{R}$. The polarization of the pump beam is modulated between $\sigma^{+}$ and $\sigma^{-}$ by a photo-elastic modulator (PEM), operated at the frequency of 84 kHz. The circularly-polarized pump pulses generate carrier spin polarization, which is then probed by the Kerr rotation of the linearly-polarized probe pulses after reflection from the sample. The Kerr rotation is measured by splitting the reflected probe beam into two beams with orthogonal polarization (using a Wollaston prism) which are then detected with a balanced photodetector and lock-in amplifier synchronized with the modulation of the pump helicity. The temporal evolution of the spin polarization is traced by varying the time delay between pump and probe pulses. 
The laser photon energy was set to 1.510~eV. The average pump (probe) power is $4$~$\mu$W ($200$~$\mu$W), while the spot size of the pump and probe beams on the sample is about 100~$\mu$m.

\section*{Author information}
\subsection*{Corresponding Authors}
 \textbf{Nataliia E. Kopteva} -- Experimentelle Physik 2, Technische Universit\"at Dortmund, 44227 Dortmund, Germany; orcid.org/0000-0003-0865-0393; Email: natalia.kopteva@tu-dortmund.de\\
 
 \textbf{Dmitri R. Yakovlev} -- Experimentelle Physik 2, Technische Universit\"at Dortmund, 44227 Dortmund, Germany; orcid.org/0000-0001-7349-2745; Email: dmitri.yakovlev@tu-dortmund.de\\
\subsection*{Authors}

 \textbf{Dennis Kudlacik} -- Experimentelle Physik 2, Technische Universit\"at Dortmund, 44227 Dortmund, Germany; orcid.org/0000-0001-5473-8383\\
 
 \textbf{Mladen Kotur} -- Experimentelle Physik 2, Technische Universit\"at Dortmund, 44227 Dortmund, Germany; orcid.org/0000-0002-2569-5051\\
 
 \textbf{Kirill V. Kavokin} -- Spin Optics Laboratory, Saint Petersburg State University, 198504 Saint Petersburg, Russia; orcid.org/0000-0002-0047-5706\\
 
 \textbf{Carolin Harkort} -- Experimentelle Physik 2, Technische Universit\"at Dortmund, 44227 Dortmund, Germany; orcid.org/0000-0003-1975-9773\\
 
 \textbf{Marek Karzel} -- Experimentelle Physik 2, Technische Universit\"at Dortmund, 44227 Dortmund, Germany; orcid.org/0000-0002-1939-5191\\
 
 \textbf{Evgeny~A.~Zhukov} -- Experimentelle Physik 2, Technische Universit\"at Dortmund, 44227 Dortmund, Germany; orcid.org/0000-0003-0695-0093   \\ 
 
 \textbf{Eiko~Evers} -- Experimentelle Physik 2, Technische Universit\"at Dortmund, 44227 Dortmund, Germany; orcid.org/0000-0002-5315-8720\\
 
 \textbf{Vasilii~V.~Belykh} -- Experimentelle Physik 2, Technische Universit\"at Dortmund, 44227 Dortmund, Germany; orcid.org/0000-0002-0032-748X \\
 
 \textbf{Manfred Bayer} -- Experimentelle Physik 2, Technische Universit\"at Dortmund, 44227 Dortmund, Germany; orcid.org/0000-0002-0893-5949\\

\subsection*{Notes}
Authors declare no competing financial interests.

\subsection*{Acknowledgements}
Samples for these studies were provided by O. Hordiichuk,  D. N. Dirin, and M. V. Kovalenko from ETH Zürich. The authors are thankful to D. S. Smirnov, V. L. Korenev, I. A. Akimov, M. O. Nestoklon,  and M. M. Glazov for fruitful discussions. We acknowledge the financial support by the Deutsche Forschungsgemeinschaft (Project YA 65/28-1, No. 527080192 associated with the SPP2196 Priority Program). KVK acknowledges the financial support by the Saint Petersburg State University through Research Grant No. 95442589.


\clearpage
\newpage
\begin{mywidetext}
\begin{center}
  \textbf{{\Large Supporting Information:}}\\
  \textbf{{\Large Optical orientation of localized electrons and holes interacting with nuclei in an {FA}$_{0.9}$Cs$_{0.1}$PbI$_{2.8}$Br$_{0.2}$ perovskite crystal}} \\
{Dennis~Kudlacik, Nataliia~E.~Kopteva, Mladen~Kotur, Dmitri~R.~Yakovlev, Kirill V. Kavokin, Carolin Harkort, Marek Karzel,  Evgeny~A.~Zhukov, Eiko~Evers, Vasilii~V.~Belykh, and Manfred~Bayer} 
\end{center}
\end{mywidetext}
\setcounter{equation}{0}
\setcounter{figure}{0}
\setcounter{table}{0}
\setcounter{page}{1}
\setcounter{section}{0}
\makeatletter
\renewcommand{\thepage}{S\arabic{page}}
\renewcommand{\theequation}{S\arabic{equation}}
\renewcommand{\thefigure}{S\arabic{figure}}
\renewcommand{\thetable}{S\arabic{table}}
\renewcommand{\thesection}{S\arabic{section}}
\renewcommand{\bibnumfmt}[1]{[S#1]}
\renewcommand{\citenumfont}[1]{S#1}


\renewcommand{\S}{\mathop{\mathcal S}}
\newcommand{\X}{\mathop{\mathcal X}}
\newcommand{\Y}{\mathop{\mathcal Y}}
\newcommand{\Z}{\mathop{\mathcal Z}}

\onecolumngrid

\section{Samples}

The studied perovskite crystals are based on the FAPbI$_{3}$ material class. FA-based perovskite exhibits a low trap density ($1.13 \times 10^{10}$~cm$^{-3}$) and a low dark carrier density ($3.9 \times 10^{9}$~cm$^{-3}$)~\cite{nazarenko2017si,zhumekenov2016si}. FAPbI$_3$ is chemically and thermally more stable compared to MAPbI$_3$ due to the decomposition of the latter to gaseous hydrogen iodide and methylammonium~\cite{nazarenko2017si}. However, pure FAPbI$_3$ suffers from structural instability originating from the large size of the FA cation, which cannot be accommodated by the inorganic perovskite framework. This instability has been successfully resolved via partial, up to 15\%, replacement of the large FA cation with smaller caesium (Cs) together with iodine (I) substitution by bromide (Br) \cite{mcmeekin2016_SI,jeon2015_SI}. As a result, the Goldschmidt tolerance factor~\cite{goldschmidt_gesetze_1926si} $t$ is tuned from 1.07 in FAPbI$_3$ closer to unity in FA$_{0.9}$Cs$_{0.1}$PbI$_{2.8}$Br$_{0.2}$, where it is 0.98~\cite{li2016si}. 

For crystal synthesis the inverse temperature crystallization technique is used~\cite{nazarenko2017si,zhumekenov2016si}. For the growth, a solution of CsI, FAI (FA being formamidinium), PbI$_2$, and PbBr$_2$, with GBL $\gamma$-butyrolactone as solvent is mixed. This solution is then filtered and slowly heated to 130$^\circ$C temperature, whereby single crystals are formed in the black phase of FA$_{0.9}$Cs$_{0.1}$PbI$_{2.8}$Br$_{0.2}$, following the reaction
\begin{equation*}
\mathrm{PbI}_2 + \mathrm{PbBr}_2 + \mathrm{FAI} + \mathrm{CsI} \underset{}{\stackrel{[GBL]}{\rightarrow}} \mathrm{FA}_{0.9}\mathrm{Cs}_{0.1}\mathrm{PbI}_{2.8}\mathrm{Br}_{0.2} + \mathrm{R}.
\end{equation*}
Afterwards the crystals are separated by filtering and drying. A typical crystal used for this study  has a size of about 2~mm. The crystallographic analysis suggests that one of the principal axis $a$, $b$, $c$ is normal to the front facet, thus pointing along the optical axis. In cubic approximation $a=b=c$. The pseudo-cubic lattice constant for hybrid organic perovskite (HOP) is around 0.63~nm~\cite{whitfield2016si}, but was not determined for this specific sample.

\section{Spectral dependence of optical orientation degree}

\begin{figure*}[htb]
\begin{center}
\includegraphics[width = 15cm]{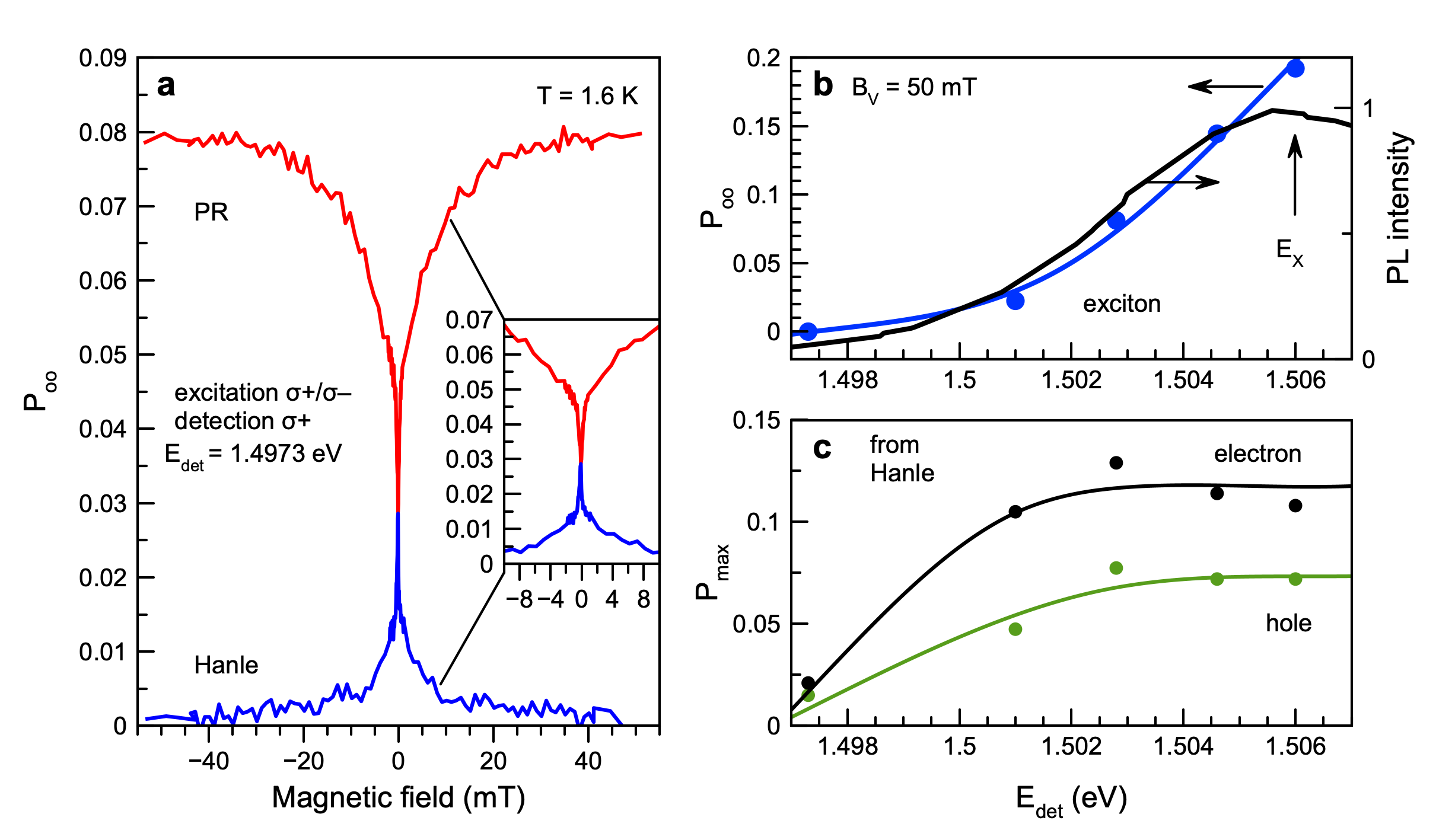}
\caption{\label{det} Spectral dependence of the optical orientation degree of excitons, electrons, and holes. (a) Hanle (blue) and PR (red) curves measured at $E_\text{det} = 1.497$~eV. The excitation helicity is modulated at $f_\text{m} = 50$\,kHz frequency and $\sigma^+$ detection is used. $T = 1.6$~K, $E_\text{exc} = 1.553$~eV, and $P = 0.01$~W/cm$^2$. (b) $P_\text{oo}$ of excitons in dependence on the detection energy measured at $B_\text{V} = 50$~mT (blue dots). Blue line is a guide to the eye. Black line corresponds to the PLE spectrum from Figure~1a, where $E_{\rm X}=1.506$~eV is the free exciton energy.  (c) Spectral dependence of the amplitudes of the electron and hole polarizations in Hanle curve.} 
\end{center}
\end{figure*}

Here we give more details on the spectral dependence of the optical orientation contributions from electrons, holes, and excitons, distinguished by the Hanle and PR measurements. It is shown in Figures~\ref{detuning}a and \ref{detuning}b of the main text, that the HWHM of the PR and Hanle curves for the electrons and holes do not depend on the excitation energy. We have checked that they are also independent of the detection energy for a fixed excitation energy. The time-resolved PL dynamics measured on the FA$_{0.9}$Cs$_{0.1}$PbI$_{2.8}$Br$_{0.2}$ crystal show that the exciton recombination, measured within the exciton lifetime of 60~ps, has its maximum at the free exciton energy of 1.506~eV and its contribution at the energy of 1.496~eV, corresponding to the PL maximum in the time-integrated measurements, is absent~\cite{XOOsi}. This is confirmed by the Hanle curves measured at 1.497~eV, Figure~\ref{det}a. One can see that the Hanle signal shown by the blue line approaches zero level in magnetic fields exceeding 40~mT, where the electron and hole optical orientations are suppressed. No contribution of the broad exciton Hanle curve is detected at this energy. One can compare that with Figure~\ref{Hanle_PRC}a, where the exciton optical orientation of 0.10 measured at 1.504~eV is well seen up to $B_{\rm V}=60$~mT. 

The spectral dependences of the optical orientation amplitudes for the excitons and carriers are given in Figures~\ref{det}b and \ref{det}c, respectively. They differ from each other: for electrons and holes they are about constant in the spectral range of $1.501-1.506$~eV with $P_\text{max,e} = 0.12$ and $P_\text{max,h}=0.06$, and decrease for 1.497~eV while being still well detectable. Contrary to that, the exciton optical orientation decreases from 0.20 at 1.506~eV down to 0.02 at 1.501~eV and vanishes at lower energies. Note, that the experimental data in Figure~\ref{det}b are measured at $B_{\rm V}=50$~mT, where only the exciton contribution is left.  The results presented in Figures~\ref{det}b and \ref{det}c show that the carrier contributions to $P_\text{oo}$ can be spectrally isolated from the exciton one without using the time-resolved techniques.  

\section{Carrier-nuclei spin interactions}

In this Section, we develop a theoretical approach to describe optical orientation of localized electrons and holes interacting with nuclear spins. The laser light $k$-vector is parallel to the $z$-axis and, therefore, $\sigma^+$ circularly polarized light orients the spins of electrons and holes along the $z$-axis. The electrons and holes, being spatially separated, have long recombination times $\tau_\text{e(h)}$, which for the general consideration valid for both types of carriers we label as $\tau $. The charge carriers can hop between localization sites (i.e. impurity centers) and their retentiuon time at a specific site is characterized by the  correlation time $\tau_\text{c}$. During the correlation time the carrier spins undergo Larmor precession around the nuclear spin fluctuations. We consider the case of intermediate correlation time, defined by the conditions $\omega_\text{L,r}\tau_\text{c} \gg 1$ and $\tau_\text{c}<\tau_\text{c,N}$  where $\langle \omega_\text{L,r}^2 \rangle$ is the mean squared frequency of the carrier spin precession in the Overhauser field of the nuclear fluctuations, and $\tau_\text{c,N}$ is the nuclear spin correlation time. In this case, the carrier spin components perpendicular to the total effective magnetic field ${\bf B}_\text{tot} = {\bf B} + {\bf B}_\text{f} $ composed of the external magnetic field $ {\bf B}$ and the effective field of the nuclear fluctuations ${\bf B}_\text{f}$ turn on average to zero because of rapid dephasing.  As a result, the mean carrier spin after $n$ hops is obtained by consequent projecting of the initial spin vector on the total fields of $n+1$ fluctuations:
\begin{equation}
\label{SE01}
\langle S_{nz} \rangle = S_0 \left \langle \frac{({\bf e}_{z} {\bf B}_\text{tot,0})({\bf B}_\text{tot,0}\cdot{\bf B}_\text{tot,1})\ldots ({\bf B}_{\text{tot},n} {\bf e}_{z}))}{B_\text{tot,0}^2 B_\text{tot,1}^2 \ldots B_{\text{tot},n} ^2}\right \rangle.
\end{equation}
Here $\langle S_{nz} \rangle$ is the average spin polarization after $n$ hops along the $z$-axis. $S_0$ is the spin polarization created by the light.  ${\bf e}_z$ is the unity vector along the $z$-axis. ${\bf B}_{\text{tot},i} = {\bf B} + {\bf B}_{\text{f},i} $ is the total magnetic field composed of the external magnetic field ${\bf B}$ and the effective field of the nuclear fluctuation ${\bf B}_{\text{f},i} $ at the $i$-th center ($i = 1\ldots n$). The statistical independence of the nuclear spin fluctuations in different localization centers allows us to factorize the equation: 
\begin{equation}
\label{SE02}
\langle S_{nz} \rangle = S_0 {\bf e}_z  \prod\limits_{i = 0}^{n} \left \langle \frac{{\bf B}_{\text{tot},i} \otimes {\bf B}_{\text{tot},i} }{B_{\text{tot},i} ^2}\right \rangle {\bf e}_{z}.
\end{equation}
The expression in the angular brackets in Eq.~\eqref{SE02} represents a random $3\times3$ matrix. Considering the general case of the external magnetic field directed at the angle $\theta$ relative to $z$ (particular cases of PR and Hanle experiments correspond to $\theta=0$ and $\theta=\pi/2$, respectively), we introduce a new coordinate system with the $z^\prime$ axis directed along the external field. Then the random matrix takes the form:
\begin{equation}
\label{SE03}
\frac{{\bf B}_{\text{tot},i} \otimes {\bf B}_{\text{tot},i}}{B_{\text{tot},i}^2} = \frac{1}{B_{\text{fx}^\prime,i}^2 + B_{\text{fy}^\prime,i}^2 + (B_{\text{fz}^\prime,i} + B_{\text{z}^\prime})^2}\left(\begin{array}{ccc} B_{\text{fx}^\prime,i}^2  & B_{\text{fx}^\prime,i}B_{\text{fy}^\prime,i} & (B_{\text{fz}^\prime,i} + B_{\text{z}^\prime})B_{\text{fx}^\prime,i} \\
B_{\text{fx}^\prime,i}B_{\text{fy}^\prime,i}   & B_{\text{fy}^\prime,i}^2 & (B_{\text{fz}^\prime,i} + B_{\text{z}^\prime})B_{\text{fy}^\prime,i} \\
 (B_{\text{fz}^\prime,i} + B_{\text{z}^\prime})B_{\text{fx}^\prime,i} &  (B_{\text{fz}^\prime,i} + B_{\text{z}^\prime})B_{\text{fy}^\prime,i} &  (B_{\text{fz}^\prime,i} + B_{\text{z}^\prime})^2\end{array}\right),
\end{equation}
where $B_{\text{fx}^\prime,i}$, $B_{\text{fy}^\prime,i}$ and $B_{\text{fz}^\prime,i}$ are the components of the nuclear fluctuation field. Since these components are not correlated in the Gaussian approximation, the averaging of the equation~\eqref{SE03} leaves only diagonal elements: 
\begin{equation}
\label{SE04}
 \left \langle \frac{{\bf B}_{\text{tot},i} \otimes {\bf B}_{\text{tot},i}}{B_{\text{tot},i}^2} \right \rangle = \left(\begin{array}{ccc} A  & 0 & 0 \\
0  & A & 0\\
0 & 0 &  1-2A \end{array}\right),
\end{equation}
\begin{equation}
\label{SE05}
A =  \left \langle \frac{B_{\text{fx}^\prime,i}^2}{B_{\text{fx}^\prime,i}^2 + B_{\text{fy}^\prime,i}^2 + (B_{\text{fz}^\prime,i} + B_{\text{z}^\prime})^2} \right \rangle =  \left \langle \frac{B_{\text{fy}^\prime,i}^2}{B_{\text{fx}^\prime,i}^2 + B_{\text{fy}^\prime,i}^2 + (B_{\text{fz}^\prime,i} + B_{\text{z}^\prime})^2} \right \rangle.
\end{equation}
By substituting Eq.~\eqref{SE04} into Eq.~\eqref{SE02}, one obtains the $z$-projection of the carrier mean spin on the $z$-axis after $n$ hops:
\begin{align}
\label{SE07}
\langle S_nz \rangle = S_0 {\bf e}_z\left(\begin{array}{ccc} A^{n+1}  & 0 & 0 \\
0  & A^{n+1} & 0\\
0 & 0 &  (1-2A)^{n+1} \end{array}\right){\bf e}_{z} =  S_0\left(\begin{array}{c} \sin \theta \\
0\\
\cos \theta \end{array}\right) \left(\begin{array}{ccc} A^{n+1}  & 0 & 0 \\
0  & A^{n+1} & 0\\
0 & 0 &  (1-2A)^{n+1} \end{array}\right) \left(\begin{array}{c} \sin \theta \\
0\\
\cos \theta \end{array}\right) =\notag \\= S_0 \left [ A^{n+1} \sin^2 \theta + (1-2A)^{n+1} \cos^2 \theta \right] .
\end{align}

Numerical calculations show that the dependence of $A$ on the external magnetic field can be approximated by an analytical expression:
\begin{equation}
\label{SE09}
A (B) \approx \frac{1}{3}\cdot \frac{1}{1 + \frac{1}{4}\frac{B^2}{\Delta_\text{N}^2}}.
\end{equation}
Here, $\Delta_\text{N}^2$ is the mean square of the projection of the nuclear spin fluctuation field $B_{f\alpha}^2$ on any direction $\alpha = x, y, z$. 

The probability that $n$ hops occurred in time $t$ is given by the Poisson distribution: 
\begin{equation}
\label{SE10}
p_{n}(t) = \frac{\langle {n} \rangle^{n}}{{n}!}\exp(-\langle {n} \rangle) = \frac{1}{{n}!}\left(\frac{t}{\tau_\text{c}}\right)^{n}\exp(-t/\tau_\text{c}).
\end{equation}
To find the average spin polarization as a function of the time elapsed after excitation, one should average Eq.~\eqref{SE07} over $n$ with the following distribution function: 
\begin{align}
\label{SE11}
\langle S_{z}(t) \rangle = S_0 \sum_{n} \left [ A^{n+1} \sin^2 \theta + (1-2A)^{n+1} \cos^2 \theta \right] \frac{1}{{n}!}\left(\frac{t}{\tau_\text{c}}\right)^{n}\exp(-t/\tau_\text{c}) =\notag \\= S_0 [A \sin^2 \theta \exp(-t[1 - A]/\tau_\text{c}) + (1-2A) \cos^2 \theta \exp(-2At/\tau_\text{c})].
\end{align}
In particular, the time dependence of the spin polarization of carriers after excitation by a short pulse of circularly polarized light in the Faraday geometry comprises a rapid (on the timescale of $\tau_\text{short}\approx 1/\sqrt{\langle\omega_\text{L,r}^2 \rangle}\ll\tau_\text{c}$) decrease from the initial value of $p_0=2S_0$ down to $p_0(1-2A)$, followed by an exponential decay with the time constant  
\begin{equation}
\label{SE12}
\tau_\text{long}=\tau_\text{c}/2A=\frac{3}{2}\tau_\text{c}\left(1 + \frac{B^2}{4\Delta_\text{N}^2}\right).
\end{equation}
If other spin relaxation mechanisms, characterized by the time $\tau_{\rm {s,0}}$ are present, the time dependence given by Eq.~\eqref{SE11} should be multiplied by $\exp{(-t/\tau_{\rm {s,0}})}$.
To get the carrier spin polarization for continuous wave excitation, one should integrate Eq.~\eqref{SE11} over the spin lifetime $\exp{(-t/\tau)}/\tau$.  This results in the following expression for the polarization recovery effect in the Faraday geometry ($\theta=0$):
\begin{equation}
\label{SE13}
\langle S_z(B_\text{F})\rangle = \frac{T_{\rm {s,0}}}{\tau}S_0 (1-2A) \left [ 1 + 2A\frac{T_{\rm {s,0}}}{\tau_\text{c}}\right ]^{-1} = \frac{T_{\rm {s,0}}}{\tau}S_0 \left(1-\frac{2T_{\rm {s,0}}+2\tau_\text{c}}{2T_{\rm {s,0}}+3\tau_\text{c}}\frac{1}{1+\frac{B_\text{F}^2}{\Delta_\text{PR}^2}}\right),
\end{equation}
where $T_{\rm {s,0}}=(\tau_{\rm {s,0}}^{-1} + \tau^{-1})^{-1}$, while for the Hanle effect measured in the Voigt geometry ($\theta=\pi/2$) we get:
\begin{equation}
\label{SE14}
\langle S_z(B_\text{V})\rangle = \frac{T_{\rm {s,0}}}{\tau}S_0 A \left [ 1 + \frac{T_{\rm {s,0}}}{\tau_\text{c}}(1 - A)\right ]^{-1} =\frac{T_{\rm {s,0}}}{\tau}S_0 \frac{\tau_\text{c}}{2T_{\rm {s,0}}+3\tau_\text{c}}\frac{1}{1+\frac{B_\text{V}^2}{\Delta_\text{H}^2}}.
\end{equation}
Here 
\begin{equation}
\label{SE15}
\Delta_\text{PR}^2=4\Delta_\text{N}^2\frac{2T_{\rm {s,0}}+3\tau_\text{c}}{3\tau_\text{c}},
\end{equation}
\begin{equation}
\label{SE16}
\Delta_\text{H}^2=4\Delta_\text{N}^2\frac{2T_{\rm {s,0}}+3\tau_\text{c}}{3T_{\rm {s,0}} + 3\tau_\text{c}}.
\end{equation}
These expressions can also be obtained from Eqs.~(18,19) of Ref.~\onlinecite{smirnov2023si} in the case of long correlation time.

Note, that equations~\eqref{SE13},\eqref{SE14} are universal for the electron and hole average spin polarization. The measured degree of optical orientation, contributed by both spin-polarized electrons and holes can be calculated as:  
\begin{equation}
\label{SE15}
P_\text{oo} = \frac{2\langle S_{z,\text{e}}(B_\text{F(V)})\rangle + 2\langle S_{z,\text{h}}(B_\text{F(V)})\rangle}{1 + 4\langle S_{z,\text{e}}(B_\text{F(V)})\rangle\cdot\langle S_{z,\text{h}}(B_\text{F(V)})\rangle} = \frac{P_\text{oo,e} + P_\text{oo,h}}{1 + P_\text{oo,e}P_\text{oo,h}}.
\end{equation}
And in case of a low degree of the carrier optical orientation it can be simplified to an additive sum of the electron and hole contributions:
\begin{equation}
\label{SE16}
P_\text{oo} = 2\langle S_{z,\text{e}}(B_\text{F(V)})\rangle + 2\langle S_{z,\text{h}}(B_\text{F(V)})\rangle = P_\text{oo,e} + P_\text{oo,h}.
\end{equation}
The final analytical equations used for fitting the experimental data are:
\begin{eqnarray}
\label{SI_PRC_NF}
P_\text{oo}(B_\text{F}) = \sum_\text{i=e,h}\frac{(3B^2_\text{F} + 4\Delta_\text{N,i}^2)P_\text{max,i}}{3B^2_\text{F} + 4\Delta_\text{N,i}^2(3 + 2T_{\rm {s,0}}/\tau_\text{c})},
\end{eqnarray}
\begin{eqnarray}
\label{SI_Hanle_NF}
P_\text{oo}(B_\text{V}) = \sum_\text{i=e,h}\frac{4\Delta_\text{N,i}^2P_\text{max,i}}{3B^2_\text{V}(1 + T_{\rm {s,0}}/\tau_\text{c}) + 4\Delta_\text{N,i}^2(3 + 2T_{\rm {s,0}}/\tau_\text{c})}.
\end{eqnarray}

\end{document}